\documentclass[prl,amsmath,twocolumn,showpacs,superscriptaddress]{revtex4}

\usepackage{graphicx}
\begin{document}

\title{ 
Infinite densities for L\'evy walks 
 }
\author{A. Rebenshtok}
\affiliation{Department of Physics, Institute of Nanotechnology and Advanced Materials, Bar-Ilan University, Ramat-Gan,
52900, Israel}
\author{S. Denisov}
\affiliation{Institute  of Physics, University of Augsburg,
Universit\"atsstrasse 1, D-86135,  Augsburg 
Germany}
\affiliation{Department for Bioinformatics, Lobachevsky State University,
Gagarin Avenue 23, 603950 Nizhny Novgorod, Russia}
\affiliation{Sumy State University, Rimsky-Korsakov Street 2, 40007 Sumy, Ukraine}
\author{P. H\"anggi}
\affiliation{Institute  of Physics, University of Augsburg,
Universit\"atsstrasse 1, D-86135,  Augsburg 
Germany}
\affiliation{Nanosystems Initiative Munich, Schellingstr, 4, D-80799 M\"unchen, Germany}
\author{E. Barkai}
\affiliation{Department of Physics, Institute of Nanotechnology and Advanced Materials, Bar-Ilan University, Ramat-Gan, 
52900, Israel}

\begin{abstract}
Motion of particles in many systems exhibits a mixture between periods
of random diffusive like events and  ballistic like motion.
In many cases, such systems exhibit strong anomalous diffusion, where low order moments
$\langle |x(t)|^q \rangle$ with $q$ below a critical value $q_c$ 
exhibit diffusive scaling while for $q>q_c$ a ballistic scaling emerges.
The mixed dynamics constitutes a theoretical
challenge since it does not fall into
a unique category of motion, e.g., the known
diffusion equations and central limit theorems fail to describe both aspects.
In this paper we resolve this problem by resorting to the concept of infinite density. 
Using the widely applicable  L\'evy walk model, 
we  find a general expression for the corresponding non-normalized 
density which is fully determined by the 
particles velocity distribution, the anomalous diffusion 
exponent $\alpha$ and the diffusion coefficient $K_\alpha$.
We  explain how infinite densities
play a central role in the description
of  dynamics of a large class of physical processes and discuss 
how they can be evaluated from experimental or numerical data. 
\end{abstract}
\maketitle

\section{Introduction}

 The trajectory of a  particle embedded in a complex or even some
seemingly  simple structures
may exhibit simultaneous modes
of motion
\cite{r_arkady_pik,Strong}. 
An example is deterministic transport of a tracer
particle in an infinite horizon ordered Lorentz billiard, 
a set of fixed circular hard scatterers arranged in a square lattice
\cite{Bou85,Zacharel,Bouchaud,Artuso,Armstead,Sanders,Courbage}.
A tracer moving freely
among a set of  scatterers
and bouncing elastically once encountering 
one of them, exhibits 
intermittent behavior with long flights where the particle moves
ballistically, separated by collision events which induce   
a diffusive like motion. A similar behavior of the tracing  particle 
can be induced by the flow acting in the phase space of chaotic Hamiltonian systems \cite{2dHam1,2dHam2}.
The key to our discussion are power law
distributed waiting times between collision events, induced by the geometry
of the scatterers \cite{Bouchaud,Levitz,BarkaiLL,Wiersma}.
 A condition for such non-Drude like dynamics
is that the radii of scatterers is smaller than half the lattice spacing 
and that the tracer is a point like particle, hence the tracer has
an infinite horizon \cite{Bouchaud,Sanders}.
   Also consider the very different type
of  motion of polymeric particles in living
cells, where sub-diffusive motion is separated by long power law
distributed  flights which induce
super-diffusion \cite{Naama}. Such systems are difficult to characterize since
they exhibit at least two modes of motion. A common tool in the analysis 
of such data is the spectrum of  exponents $q \nu(q)$ \cite{Strong}.  
One measures the $q$-th moment of the motion, for particles starting
on a common origin
\begin{equation}
\langle |x(t)|^q \rangle \sim t^{ q \nu(q)}
\label{eqInt01}
\end{equation} 
with $q>0$. Here, for sake of simplicity
 we consider the one dimensional case and avoid other
possible time dependencies like  a logarithmic increase of moments with time. 
Scale invariant transport implies that $\nu(q)$ is a constant independent of
$q$, for example,
for Brownian motion $\nu(q) =1/2$. In many fields of science 
one finds
that $\nu(q)$ is a non-linear function of $q$, the term strong anomalous
diffusion is often used in this context \cite{Strong}. Surprisingly, in many cases 
the continuous 
spectrum  $q \nu(q)$ exhibits a bi-linear scaling (see details below).
 Examples for
this piecewise linear behavior of $q \nu(q)$ include
nonlinear dynamical systems 
 \cite{Strong,Artuso,Armstead,Sanders,Courbage},
stochastic models with quenched and annealed disorder,
 in particular,  the L\'evy walk \cite{Andersen,GL,Zabu1,r_vezzani,Burioni,Bernabo} and sand pile models \cite{Sandpile}. 
Recent experiments 
on  the  active transport of polymers in the cell \cite{Naama},
theoretical investigation of the  momentum \cite{KesslerPRL}
and the spatial \cite{DechantPRL} spreading  of cold atoms in optical lattices
and flows in porous media \cite{Anna}
further confirmed the generality of strong anomalous
diffusion of the bi-linear type. 
 An  example
is a diffusive scaling $q \nu(q) = q/2$ below a certain value
of $q_c>0$ and a ballistic scaling $q \nu(q)=q - q_c/2$ otherwise. This is 
an example of a motion which is neither purely diffusive nor
ballistic.
 
 While the mechanisms leading to strong anomalous diffusion
could be vast, even
the stochastic treatment of such systems is not well understood. 
For example, diffusion equations, either normal or fractional 
\cite{Review},
stochastic frameworks like fractional Brownian motion or generalized
Langevin equations \cite{Goy}  and the
Gauss-L\'evy  central limit
theorem \cite{Bouchaud,levy} fail to describe this phenomena.
We recently investigated the L\'evy walk  process 
\cite{Shles,KBS,Klafter,ZK,Cristadoro},
a
well known stochastic model of many natural 
behaviors which exhibits bi-linear strong anomalous diffusion. 
Examining four
 special cases we found that a non-normalizable
density, with a diverging area underneath it,  describes these processes 
\cite{gang4}.
 This unusual infinite density,
describes the ballistic  aspects of strong anomalous diffusion.
Here we find a general formula for the infinite density which, 
as we show below, is complementary
to the Gauss and L\'evy distributions. 
The latter are the mathematical basis of many diffusive phenomena, and here
we promote the idea that the
infinite  density concept  is  a rather general
tool statistically characterizing  the ballistic trait of the motion.  
We show how these unnormalized distributions emerge from a basic model
with wide applications, 
thus, possibly this overlooked measure may become an important
tool.

 In mathematics the concept of a non-normalized 
 infinite density was thoroughly investigated
\cite{Thaler1,Aaronson,Thaler}
while in physics this idea has gained interest only recently (see below). 
Conservation of matter implies that 
the number of particles in the system is fixed, thus naturally
one may normalize the probability density describing a packet of non-interacting
diffusing particles $P(x,t)$
to unity 
$\int_{-\infty} ^\infty P(x,t) {\rm d} x=1$. Hence dynamical and
equilibrium properties  of most  physical
systems are described by densities which are normalized, for example,
a Boltzmann-Gibbs state in thermal equilibrium, or solutions
of Boltzmann or Fokker-Planck equations with normalization 
conserving boundary conditions (no absorbing boundaries).  However, in 
some cases a closer distinction must be made, namely a distinction based
on the observable of interest. 
Probably the most common averaged observables are
the moments  $\langle |x|^q \rangle$ of a process $x(t)$, so in this example
the observable is $|x(t)|^q$ and we will distinguish between high order moments
$q>q_c$ and low order ones.  In our case we treat a system
with mixed dynamics, and show that the low order moments are described
by the standard machinery of non-equilibrium statistical physics (fractional
diffusion equations and L\'evy central limit theorem) but the high order
moments, which represent the ballistic elements of the process, are
described by an infinite density. In this sense the infinite
density is complementary to the central limit theorem \cite{gang4}.
 We find a general
formula for this infinite density relating it
to the velocity distribution of the particles, the anomalous diffusion
exponent $\alpha$ and the diffusion constant $K_\alpha$.
Our work shows how the unnormalized state 
emerges from the norm conserving dynamics of the L\'evy walk.  
Previous work
\cite{KesslerPRL,KorabelPesin,Miya01,Akimotoprl,KorabelINF,KorabelNum,LutzNature,Holz} on applications of the infinite density in physics, 
  dealt with bounded systems which attain an equilibrium.
 For example
the momentum distribution of cold atoms where the Gibbs-measure is finite 
the infinite density describes the large rare
 fluctuations of the kinetic energy \cite{KesslerPRL,Holz}.
Or intermittent maps  with unstable fixed points,
e.g., the  Pomeau-Manneville transformation on the unit interval
\cite{KorabelPesin,Akimotoprl}.
Here 
we consider systems that are not in equilibrium, and the dynamics is unbounded, showing that the potential applications of infinite densities are vast.

\section{L\'evy walk model}

In the L\'evy walk process the trajectory of the particle $x(t)$  consists
of epochs of ballistic travel separated by a set of collisions
which alter its velocity \cite{Bouchaud,Review,Shles,Klafter,ZK}.
 At time $t=0$ the particle is on the origin.
Its initial velocity $-\infty< v_0<\infty$
 is random and drawn from  a probability density
function  (PDF) $F(v)$, whose moments are all non-diverging \cite{LWbasic}. The particle travels with a constant
speed for a random  duration $\tau_1>0$ 
whose PDF is $\psi(\tau)$.
The particle's displacement in this first sojourn time is 
$v_0 \tau_1$. At time $t_1=\tau_1$ we draw a new velocity
$v_1$ and
a waiting time 
$\tau_2$
 from the corresponding
PDFs $F(v)$ and  $\psi(\tau)$. The second displacement is $v_1 \tau_2$.
This process is renewed. The waiting times $\{ \tau_i\}$ ($i=1,2,\cdots$)
 and the velocities
$\{ v_j \}$ ($j=0,1,\cdots$) 
are  mutually independent identically distributed random variables.
 The points on the time axis $t_N =\sum_{i=1}^N \tau_i$ are the collision
times, when the particle switches its velocity.  
The position of the particle at time $t$ is
\begin{equation}
x(t) = \sum_{j=1} ^N v_{j-1} \tau_j + v_N \tau_b\:.
\label{eqModel00}
\end{equation}
Here $N$ is the random number of collisions or renewals in the time
interval $(0,t)$.  The time interval
$\tau_b = t-t_N$ is called the backward recurrence
time \cite{GL}.
 The last term in Eq. (\ref{eqModel00}) describes the motion between
the last collision event and the measurement time $t$  
\begin{equation}
t = \sum_{j=1} ^N \tau_j + \tau_b\:.
\label{eqModel00a}
\end{equation}
A model where $N$ is fixed and $t$ is random, namely, a process stopped
after $N$ collisions 
(so $\tau_b=0$) is the classical problem of summation of independent
identically distributed random variables for which the
L\'evy-Gauss central limit theorem
applies \cite{levy}.  

As already mentioned, we assume that $F(v)=F(-v)$ hence from symmetry the 
density of particles $P(x,t)$ is also symmetric since all
particles start on the origin. We also assume that the even
 moments
of $F(v)$ are finite, hence the  tail of $F(v)$ decays
faster than any power law. The PDF of waiting times is given
in the long time limit by
\begin{equation}
\psi(\tau) \sim { A \over | \Gamma(-\alpha)|} \tau^{ - 1 - \alpha}\:,
\label{eqModel00b}
\end{equation}
where $A>0$ and 
\begin{equation}
1<\alpha< 2\:.
\label{eqModel00c}
\end{equation}
As discussed briefly in the summary,
our main finding, that a non-normalized infinite
density describes the density of particles $P(x,t)$, is found
also for $\alpha\ge 2$, but for simplicity of presentation we
consider only a limited interval for $\alpha$. The case
$1<\alpha<2$ corresponds to enhanced diffusion $\langle x^2 \rangle \propto
 t^{3 -\alpha}$ which is faster than normal diffusion but slower than ballistic
\cite{Klafter}.
  The regime $\alpha<1$, 
is called the ballistic phase of the motion
 since  then $\langle x^2 \rangle \propto t^2$. 
For that case and
for a general class of the
velocity PDF, $F(v)$, 
the  PDF $P(x,t)$ 
is given by a formula
in \cite{Rebenshtok,Rebenshtok1}.
The case $\alpha>2$ corresponds to normal diffusion in the mean square
displacement sense $\langle x^2 \rangle \propto t$.

A convenient tool is 
the Laplace transform of the waiting times PDF  denoted
\begin{equation}
\hat{\psi}(u) = \int_0 ^\infty \exp( - u \tau) \psi(\tau) {\rm d} \tau\:. 
\label{eqMod01}
\end{equation} 
For a power law distributed 
waiting time under investigation, i.e., $1<\alpha<2$,
and for small $u$ the following expansion holds
\cite{Review,ZK,GL,Carry}
\begin{equation}
\hat{\psi}(u) = 1 -  \langle \tau \rangle u+ A u^{\alpha} + \cdots
\label{eqMod02}
\end{equation}
where $\langle \tau \rangle = \int_0 ^\infty \tau \psi(\tau) {\rm d} \tau$ 
is the averaged waiting time. 

\section{Some background on the L\'evy walk}

The position of the particle is rewritten
as
\begin{equation}
x= \sum_{i=1}^N  \chi_i + \chi^* 
\label{eqLW01}
\end{equation}
where $\chi_i = v_{i-1} \tau_i$ are the flight lengths and $\chi^{*} = v_N \tau_b$.
Since the velocity distribution is narrow and
symmetric,
 the PDF of flight lengths 
$q(\chi)$
is also  symmetric $q(\chi) = q(-\chi)$. 
 Since the durations of the flights are power law
distributed we have 
\begin{equation}
q (\chi) \propto |\chi|^{ -(1 + \alpha)}
\label{eqLW02}
\end{equation}
for large $|\chi|$. In the regime $0<\alpha<2$ the variance of jump length
is infinite hence the Gaussian central limit theorem does not apply.
In the regime $1<\alpha<2$ the average waiting time $\langle \tau \rangle$ is
finite.  Neglecting  fluctuations, the number of flights is
 $N = t / \langle \tau \rangle$ when $t$ is long. 
Hence in this over-simplified  picture we are dealing with a 
problem of a sum of $N$ independent identically distributed random variables
with a common PDF $q(\chi)$ with a diverging variance. In this 
picture
the last jump $\chi^*$ is negligible. 
  Thus, one may argue that $x$ is described by
L\'evy's  generalized central limit theorem. This means that the PDF
of $x$ is expected to be a symmetric L\'evy stable law \cite{Kotulski}
(see details below).
However, 
such a treatment ignores the correlations between flight durations
and flight lengths and the fluctuations of $N$. In fact the finite speed of the particles implies
that jumps much larger than the typical velocity times the
 measurement time $t$
are impossible. Thus, the mean square displacement and higher order moments
always increase slower than ballistic, for example
 $\langle x^2 \rangle < \mbox{const}\ t^2$. In contrast,
a sum of independent identically distributed random variables with an infinite
variance, often called a L\'evy flight, yields a diverging mean
square displacement $\langle x^2 \rangle=\infty$. For that reason L\'evy walks,
which take into consideration the finite velocity,
 are considered more physical
if compared with L\'evy flights \cite{Strong,Klafter}.  
 Since both  the Gauss and the  L\'evy central limit theorems break 
down in the description of moments
of L\'evy walks, more specifically in the tails of the PDF of $x$,
it is natural to ask if there exists a mathematical replacement for these
widely applied theorems.  
 
 A more serious treatment of the L\'evy walk is 
given in terms of the Montroll-Weiss equation. Let $P(x,t)$ be the PDF of the position of particles at time $t$
for particles starting at the origin. From the mentioned
symmetry of the process
$P(x,t) = P(-x,t)$, $P(x,t)|_{t=0} = \delta(x)$
 so clearly all odd moments of $P(x,t)$ are
zero. 
 We define the Fourier-Laplace
transform 
\begin{equation}
P(k,u) = \int_{-\infty} ^\infty {\rm d} x \int_0 ^\infty {\rm d} t \exp( i k x - u t) P(x,t)\:. 
\label{eqLW03}
\end{equation} 
The Montroll-Weiss equation gives
the relation between the distributions of the model parameters
namely  velocities and waiting times
with $P(k,u)$ \cite{KBS,ZK,Carry} 
\begin{equation}
P(k,u) = \left\langle { 1 - \hat{\psi} (u - i k v ) \over u - i k v } \right\rangle
{ 1 \over 1 - \left\langle \hat{\psi} \left(u - i kv \right)\right\rangle} \:.
\label{eqLW04}
\end{equation}
Here the averages are with respect to the velocity distribution
$\langle ... \rangle = \int_{-\infty} ^\infty {\rm d} v \cdots F(v)$. The
derivation of this  classical result is provided in an Appendix. 
Note that in the original work of Montroll and Weiss a decoupled random walk
was considered and the origin of Eq. 
(\ref{eqLW04}) can be traced to the work
of Scher and Lax \cite{Lax}
 and that of Shlesinger, West and Klafter \cite{Shles} 
(see also \cite{KBS,ZK,Carry,pathways,Kutner,Marcin}). 
Examples \cite{Klafter,Bouchaud,Review}
 for physical processes described by the L\'evy
walk are certain non-linear dynamical systems 
\cite{Artuso,Shles,
Strong,Solomon},
polymer dynamics \cite{Ott},
blinking quantum dots \cite{Jung,Margprl,Marg,Hoog},
cold atoms diffusing in optical lattices \cite{Sagi,KesBarPRL}, 
intermittent search strategies \cite{Benichou},
dynamics of perturbations in many body Hamiltonian systems \cite{Zabu,Figueiredo,Spohn},
and particle dynamics in plasmas \cite{Zimbardo}.

 In what follows we naively expect that one can reconstruct 
the normalized  density of particles in the long time limit from the exact expressions for the moments of the process.  
Specifically, we obtain the exact expression for the long time limit of the integer moments of the 
L\'evy walk process $\langle x^{2 n } (t) \rangle$ with $n=1,2, ..$, where the case $n=0$ is trivial 
since $\langle x^0 \rangle = \int_{-\infty} ^\infty P(x,t) {\rm d} x =1$, and then we use the moments to consruct a series
equivalent to the Fourier transform of the density
\begin{equation} 
 \langle \exp(i k x)\rangle= \int_{-\infty} ^\infty e^{ i k x} P(x,t) {\rm d} x=
1 + \sum_{n=1} ^{\infty} {\langle  (i k x)^{2 n} \rangle  \over (2 n)!}. 
\label{eqplan}
\end{equation} 
Luckily, we can evaluate analytically the sum for the model under consideration.
Then we perform
the inverse Fourier transform of the such obtained function.
 One then naively expects to get
the long time limit of the density $P(x,t)$ since we use
the long time limit of the moments. It turns out that this procedure
yields a density which is {\it not} normalizable 
(for reasons which will become clear later). 
However, while the solution is not normalizable, it still describes
the density of particles $P(x,t)$ in ways which will hopefully
 become  more transparent
to the reader by the end of the manuscript.

\begin{widetext}
\section{The Moments}

 To obtain the long time behavior
of  spatial moments $\langle x^{2n}(t)\rangle$
we first find the Laplace transforms $\langle x^{2 n} (u) \rangle$.
Here, as mentioned, 
 odd moments are zero due to the assumed symmetry $F(v)=F(-v)$. 
For that aim we use the well known expansion
\begin{equation}
P(k,u) = \int_{-\infty} ^\infty e^{ i k x } P(x,u) {\rm d} x = \int_{-\infty} ^\infty \left[ 1 + \sum_{n=1} ^\infty { (i k x)^{2 n} \over (2 n)!} \right] P(x,u) {\rm d} x=  { 1 \over u} + \sum_{n=1} ^\infty { (i k )^{2 n}  \langle x^{2 n} (u) \rangle \over (2 n)! }\:.
\label{eqM01}
\end{equation}
Expanding the numerator and denominator of the
 Montroll-Weiss equation,  
Eq. (\ref{eqLW04})
using the expansion Eq. 
(\ref{eqMod02}), 
to the leading order in the small
parameter $u^{\alpha -1}$, while keeping the  ratio $k/u$ fixed, we get
\begin{equation}
P(k,u) \sim { 1 - \tilde{A} u^{\alpha -1} \langle
\left( 1 -  {i k v \over u} \right)^{\alpha -1} \rangle \over u \left[ 1 - \tilde{A} u^{\alpha -1} \langle \left( 1 -  { ik v \over u} \right)^\alpha \rangle \right]}\:
\label{eqM02}
\end{equation}
with $\tilde{A}= A / \langle \tau \rangle$. 
As is well known, 
such small $u$ expansions
correspond to the long time limit \cite{Review,GL}.  
Further expanding the denominator to find the first non-trivial 
term we obtain
\begin{equation}
P(k,u) \sim { 1 \over u} \left\{ 1 - \tilde{A} u^{\alpha -1} \left[ \left\langle \left( 1 -  { i k v \over u} \right)^{\alpha -1}\right\rangle - \left\langle \left( 1 -  { i k v \over u} \right)^\alpha \right\rangle \right] + \cdots \right\}\:. 
\label{eqM03}
\end{equation}
The leading $(1/u)$ term is obviously the normalization condition
$P(k,u)|_{k=0} = 1/u$.
In this expansion we included terms of the  order $u^{\alpha -1}$, while
higher order
terms,
which are found from further expansion
of the  denominator in Eq. (\ref{eqM02}), but also non-universal terms
which stem from the expansion of 
$\hat{\psi}(u)$, 
Eq. 
(\ref{eqMod02}),
to orders greater than $u^\alpha$, are neglected.  
We Taylor expand Eq. (\ref{eqM03}) in $k/u$, 
using the  series expansion
\begin{equation}
\left( 1 - \epsilon\right)^{\alpha -1} - \left(1-\epsilon\right)^\alpha = -\sum_{m=1} ^\infty { (-\alpha)_m \epsilon^m \over \alpha (m-1)! } \:,
\label{eqM04}
\end{equation}
where $(a)_m= \Gamma(a + m)/\Gamma(a)= a(a+1) \cdots(a + m-1)$ is the Pochhammer
symbol.
 Averaging over velocities
we find
\begin{equation}
P(k,u) \sim { 1 \over u} - \tilde{A} \sum_{n=1} ^\infty { ( 2 n) (-\alpha)_{2 n} (-1)^n \over (2 n)! (-\alpha) } u^{\alpha - 2 - 2 n} \langle v^{2n} \rangle k^{2 n}\:
\label{eqM05}
\end{equation}
with $\langle v^{2 n} \rangle = \int_{-\infty} ^\infty v^{ 2n} F(v) {\rm d} v$.
Comparing with  Eq. (\ref{eqM01}) 
yields in the small $u$ limit 
\begin{equation}
\langle x^{ 2 n} (u) \rangle \sim {\tilde{A} \over \alpha} (-\alpha)_{2 n} (2 n) \langle v^{2 n} \rangle u^{\alpha - 2 - 2 n}
\label{eqM06}
\end{equation} 
which is valid for $1<\alpha<2$ and $n=1,2,\cdots$.
Using the Laplace pair $u^{ \alpha - 2 - 2 n} \longleftrightarrow t^{2 n + 1 - \alpha} / \Gamma( 2 n + 2 - \alpha)$ we find the long time limit
of 
the even spatial moments
\begin{equation}
\langle x^{2 n}(t) \rangle \sim B { 2 n \over (2 n -\alpha) ( 2 n + 1 - \alpha) } \langle v^{2n} \rangle t^{ 2n + 1 - \alpha} \:,
\label{eq01}
\end{equation}
with $B = A / [ | \Gamma(1 - \alpha)| \langle \tau \rangle]$. 
As well known, the process exhibits super-diffusion $\langle x^2 \rangle\propto
t^{ 3 - \alpha}$. 
We now use the asymptotic result for $\langle x^{2 n} (t) \rangle$
to find the infinite
density of the L\'evy walk process. 

\section{The infinite density, $1<\alpha < 2$}

The density of particles $P(x,t)$ and its Fourier
transform $P(k,t)$ are defined according to 
\begin{equation} 
P(k,t) = \int_{-\infty} ^\infty P(x,t) e^{ i k x} {\rm d} x, \ \ \ \
P(x,t) = {1 \over  2 \pi} \int_{-\infty} ^\infty P(k,t) e^{ - i kx} {\rm d} k\:.
\label{eq03}
\end{equation}
We Taylor expand $P(k,t)$ as in Eq. 
(\ref{eqplan}), yielding
\begin{equation}
P(k,t) = 1 + \sum_{n=1} ^\infty { (i k )^{2 n} \langle x^{2 n} (t) \rangle \over (2 n)! } \:.
\label{eq04}
\end{equation} 
Clearly, we have $P(k,t)|_{k=0} = 1$ which is the normalization
condition, namely, for any finite $t$ the density of particles $P(x,t)$ is
normalized to unity. 
We next insert the exact long time expressions for the moments, 
Eq. (\ref{eq01}), 
in the series Eq. (\ref{eq04}), using 
$\langle v^{2 n} \rangle = \int_{-\infty} ^\infty v^{2 n} F(v) {\rm d} v$ 
to define
\begin{equation}
P_A ( k, t) \equiv 1 + { B \over t^{\alpha -1} } \sum_{ n=1} ^\infty \int_{-\infty} ^\infty {\rm d} v F(v) { 2 n (i k v t)^{2 n}   \over (2 n)! ( 2 n- \alpha) (2n + 1 - \alpha) } \:.
\label{eq05}
\end{equation}
Here, the subscript $A$ denotes an asymptotic expression in the sense
that we have used the long time limit of the spatial moments. 
It is convenient to define
\begin{equation}
\tilde{G}_\alpha (y) = \sum_{n=1} ^\infty { (-1)^n y^{2 n} \over
(2 n -1)! (2 n - \alpha) (2 n + 1 - \alpha) } 
\label{eq06}
\end{equation}
hence
\begin{equation}
P_A(k,t) = 1 + { B \over t^{\alpha-1} } \int_{-\infty} ^\infty 
{\rm d} v F(v) \tilde{G}_\alpha ( k v t) \:.
\label{eq07}
\end{equation}
It is easy to validate the following identity
\begin{equation}
\tilde{G}_\alpha (y) = \alpha \tilde{B}_\alpha(y) - (\alpha -1) \tilde{B}_{\alpha -1} (y)  \:,
\label{eq08}
\end{equation}
where
\begin{equation}
\tilde{B}_\alpha (y) = \sum_{n=1} ^\infty { (-1)^n y^{ 2 n} \over (2 n)! (2 n  -\alpha)}\:.
\label{eq09}
\end{equation}
With the expansion, $\cos(y) = \sum_{n=0} ^\infty (-1)^n y^{2 n} / (2n)!$,
it is also easy to see that
\begin{equation}
\tilde{B}_\alpha (y) = \int_0 ^1 { \cos( \omega y) - 1 \over \omega^{1 + \alpha} } {\rm d} \omega\:. 
\label{eq10}
\end{equation}
Using Mathematica this function is expressed in terms of the generalized
hypergeometric function 
\begin{equation}
\tilde{B}_\alpha(y)= \left\{ 1 - {}_1 F_2 \left[-{\alpha\over 2};{1\over 2},{2-\alpha\over 2};-\left({y\over 2}\right)^2\right]\right\}/\alpha\:. 
\label{eq10A}
\end{equation}
For $y>>1$ we find  $\tilde{B}_\alpha (y) \propto y^\alpha$,
thus the non-analytical behavior of $\hat{\psi}(u)$ for small
argument $u$, appears in $\tilde{B}_\alpha(y)$ when $y$ is large.

 For the inverse Fourier transform we obtain
\begin{equation}
P_A(x,t) = { 1 \over 2 \pi} \int_{-\infty} ^\infty e^{ - i k x} P_A(k, t) {\rm d} k \:.
\label{eq11}
\end{equation}
For that aim we investigate 
\begin{equation}
B_\alpha(x,v t) \equiv { 1 \over 2 \pi} \int_{-\infty} ^\infty {\rm d} k e^{ - i k x} \tilde{B}_\alpha ( k v t) \:.
\label{eq12}
\end{equation}
Using Eq. (\ref{eq10}) we have 
\begin{equation}
B_\alpha(x, v t) = { 1 \over 2 \pi} \int_{-\infty} ^\infty e^{ - i k x} {\rm d} k 
\int_0 ^1 { \cos ( \omega k v t ) - 1 \over \omega^{1 + \alpha} } {\rm d} \omega\:.
\label{eq13}
\end{equation}
The Fourier pair of $\cos(k y)$ is $[\delta(x-y) + \delta(x+y)]/2$,
hence for $x\neq 0$ 
\begin{equation}
B_\alpha( x , v t) = \left\{
\begin{array}{c c}   { 1 \over 2} { ( |v| t)^\alpha \over |x|^{1 + \alpha} }, \  &  |x|< |v|t \\
0\ \quad\quad\:\:, \  &  \;|x|> |v| t\:.
\end{array}
\right.
\label{eq14}
\end{equation}
Note that this function is not integrable. Mathematically the integral
formula 
of Fourier transform Eq. 
(\ref{eq11}) is valid for Lebesgue integrable functions (called $L^{1}$)
while we are dealing with a distribution.
 We insert Eq. (\ref{eq07}) in Eq.  (\ref{eq11}) using the definition  
Eq. (\ref{eq08}), i.e.,
\begin{equation}
P_A (x ,t) = \int_{-\infty} ^\infty {\rm d} k { e^{ - i k x} \over 2 \pi} \left\{ 1 + { B \over t^{\alpha -1} } \int_{-\infty} ^\infty {\rm d} v F(v) \left[ 
\alpha \tilde{B}_\alpha \left(k v t\right) - \left(\alpha -1\right) \tilde{B}_{\alpha-1} \left( k v t\right) \right]\right\}. 
\label{eq15}
\end{equation}
With the inverse Fourier formula Eq. (\ref{eq13}) the
$k$-integration yields
\begin{equation}
P_A (x, t) = \delta (x) + { B \over t^{\alpha -1} } \int_{-\infty} ^\infty
{\rm d} v F(v) \left[ \alpha B_\alpha \left( x, v t \right) - \left( \alpha -1\right) B_{\alpha -1} \left( x , v t\right) \right]. 
\label{eq16}
\end{equation}
The $\delta(\cdots)$ term is clearly arising from the normalization condition,
namely the $1$ in Eq. (\ref{eq15}). 
Hence using Eq. (\ref{eq14}), and the symmetries $F(v)=F(-v)$ and
$B_\alpha (x, v t) = B_\alpha (x, - v t)$ 
we find
\begin{equation}
P_{A} (x,t) =  { B \over t^{\alpha }} \int_{ |x|/t} ^\infty {\rm d}v F(v) 
\left[ \alpha { |v|^\alpha  \over |x/t|^{1 + \alpha} } -
\left( \alpha -1 \right) { |v|^{ \alpha -1}  \over |x/t|^\alpha }\right]\:,
\label{eq17}
\end{equation}
\end{widetext}
 for $x \neq 0$.
If we take $t$ to be large though finite, 
the function $P_A (x,t)$ is not normalizable, since for small
non-zero $|x|$ we get $P_A(x,t) \propto |x|^{ - (1 +\alpha)}$ and thus
 the spatial
 integral
over $P_A(x,t)$ diverges. Since the number of particles is conserved
in the underlying process, $P(x,t)$ is  normalized to unity.
One may therefore conclude that the non-normalized state
$P_A(x,t)$ does not describe the density of particles and hence
does not describe physical reality. 
This oversimplified point of view turns out to be wrong as
we proceed to show in the next section.
We now define the infinite density. 

On right hand side of  Eq. (\ref{eq17}), 
$x/t$ enters as a scaling variable, which we denote as
\begin{equation}
\overline{v} = {\int_0 ^t v(t) {\rm d}t \over t} = { x \over t}\:.
\label{eq18}
\end{equation}
Clearly $\overline{v}$ is the time average of the velocity and
as usual we consider asymptotic long times. 
The scaled variable $\overline{v}$ 
has a non-trivial density, and it describes the ballistic scaling
of the process $x \propto t$. As we discuss below,  there is not a unique
scaling in the model, and the underlying process follows also
a L\'evy scaling $x \propto t^{ 1/\alpha}$, which is a typical
scaling for  anomalous
 diffusion. These two types of scalings
are in turn  related to the bi-scaling of the moments,
found in many systems (strong anomalous diffusion).  
For now we define the infinite  density 
\begin{equation}
{\cal I} \left( \overline{v} \right) = t^\alpha P_A (x,t)
\label{eq19}
\end{equation}
In view of Eq. (\ref{eq17})
we find our main formula
\begin{equation}
{\cal I} ( \overline{v} ) = B \left[ { \alpha {\cal F}_\alpha\left( | \overline{v}| \right)  \over |\overline{v}|^{ 1 + \alpha} } - { \left( \alpha -1\right) {\cal F}_{\alpha -1} \left( |\overline{v} | \right) \over |\overline{v}|^\alpha} \right]\:,
\label{eq20}
\end{equation}
where
\begin{equation}
{\cal F}_\alpha (\overline{v} ) = \int_{ |\overline{v}|} ^\infty {\rm d}v\, v^\alpha  F(v) \:.
\label{eq21}
\end{equation}
We will soon relate the infinite density ${\cal I}(\overline{v})$,
 with the normalized L\'evy walk  probability density 
$P(x,t)$
while the definition  Eq. (\ref{eq19}) relates ${\cal I}(\overline{v})$
with $P_A(x,t)$.

{\bf Remark:} In our previous publication 
\cite{gang4} we called ${\cal I}(\overline{v})$ an infinite {\em covariant}
 density since the
 transformation of both space  $x \to c x$ and time $t \to c t$ leaves 
${\cal I}(\overline{v})/c^\alpha$ unchanged. In mathematics
infinite invariant densities usually reflect solutions which are
invariant under time shift only (steady states), e.g., for maps
with an unstable fixed point. To avoid jargon
we will call ${\cal I}(\overline{v})$ an infinite 
 density 
meaning a non-normalized density.

\subsection{Relation between ${\cal I} (\overline{v})$
and the anomalous diffusion coefficient $K_\alpha$}

In the limit $\overline{v} \to 0$, Eqs. (\ref{eq20},\ref{eq21}) give
\begin{equation}
{\cal I}\left( \overline{v} \right) \sim { B \alpha \langle |v|^\alpha \rangle \over 2} { 1 \over |\overline{v}|^{1 + \alpha}}\:.
\label{eqAs01}
\end{equation} 
This small $\overline{v}$ behavior implies that the integral
over ${\cal I}( \overline{v})$ diverges, hence the name infinite density.
One may rewrite Eq. (\ref{eqAs01}) in terms of the anomalous diffusion
constant $K_\alpha$ using
\begin{equation} 
{ B \alpha \langle |v|^\alpha \rangle \over 2} = K_\alpha c_\alpha
\label{eqAs02}
\end{equation} 
with 
\begin{equation}
K_\alpha = {A \over \langle\tau\rangle}
 \langle |v|^\alpha \rangle \left| \cos { \pi \alpha \over 2}\right|
\label{eqAs03}
\end{equation} 
and
$c_\alpha=\Gamma(1 + \alpha) \sin \left( { \pi \alpha \over 2} \right) / \pi$.
The constant $K_\alpha$ can be understood as a generalization of the standard
diffusion constant in the framework of the fractional Fokker-Planck equation, $
\partial P(x,t)/\partial t = 
K_{\alpha} \bigtriangledown^{\alpha} P(x,t)$ \cite{Review} (note that this 
equation addresses the central L\'{e}vy-like part of the PDF, $P_{\rm cen} (x,t)$, only,
see Sec.~``The L\'{e}vy scaling regime'' for a more detailed discussion). Briefly, 
it describes the width of the density field $P(x,t)$,
so it is a measurable quantity.

Further on, we find
\begin{equation}
{\cal I}\left( \overline{v} \right) \sim { c_\alpha K_\alpha \over |\overline{v}|^{1 + \alpha}}\:.
\label{eqAs04}
\end{equation} 
This relation provides the connection between the diffusive properties
of the system and the infinite density. In principle one may record in the laboratory
the spreading of the particles, then estimate  the exponent
$\alpha$ and $K_\alpha$ by observing the center part of the packet, and
afterward 
predict the small $\overline{v}$
behavior of the infinite density. 
The relation given by Eq. (\ref{eqAs04}) is general in the sense
that it does not depend on the particular form of $F(v)$.

\subsection{Asymptotic behavior of ${\cal I}(\overline{v})$}

In the opposite limit of large $\overline{v}$ we find
\begin{equation}
{\cal I}\left( \overline{v} \right) \sim B { 1 - Q\left( \overline{v} \right) \over |\overline{v} | }
\label{eqlv01}
\end{equation}
where $Q(\overline{v}) = \int_{-\infty} ^{\overline{v}} F(v) {\rm d} v$ is 
the cumulative velocity distribution 
obeying,
 $1 - Q(\overline{v})\to 0$,
as $\overline{v} \to \infty$. 
To find this result  we first integrate by parts  
Eq. (\ref{eq21})
\begin{equation}
{\cal F}_\alpha \left( \overline{v} \right) = \overline{v}^\alpha
\left[ 1 - Q(\overline{v})\right] 
+ \alpha \int_{\overline{v}} ^\infty v^{\alpha -1} \left[ 1 - Q(v)\right] {\rm d} v 
\label{eqlv02}
\end{equation}
where we used $F(v) = {\rm d} Q(v)/ {\rm d} v$, $\lim_{v \to \infty} v^\alpha[1- Q(v)] = 0$ and $\overline{v}>0$. 
In the limit of large $\overline{v}$ we may omit the second term
on the right hand side of Eq. (\ref{eqlv02}) if it is much smaller
than the first. In that case the following condition must hold
\begin{equation}
\lim_{\overline{v} \to \infty} { \alpha \int_{\overline{v}} ^\infty v^{\alpha -1} [ 1 - Q(v) ] {\rm d} v \over \overline{v}^\alpha [ 1 - Q(\overline{v})] } =0\:.
\label{eqlv3}
\end{equation}
With L'Hospital's rule we obtain the condition
\begin{equation} 
\lim_{\overline{v} \to \infty} { 
 1 - Q ( \overline{v} )\over 
\overline{v} F(\overline{v}) 
 } = 0\:.
\label{eqlv4}
\end{equation} 
Hence, if $1- Q(\overline{v})$ approaches zero faster than a power law
 for large $\overline{v}$ the condition is met 
and
\begin{equation}
{\cal F}_\alpha \left( \overline{v} \right) \sim \overline{v}^\alpha \left[ 1 - Q(\overline{v}) \right]\:.
\label{eqlv05}
\end{equation}
For example, it is easy to check this equation if $1 - Q(v)= c \exp( - v)$ for
a certain $v> v_0$ and $\overline{v}>v_0$. The equation is not
valid for a power law behavior $1- Q(v) \sim v^{ - (1 + \nu)} $ 
with $\nu>0$,
which is hardly surprising since we assume all along that moments
of $F(v)$ are either zero or finite, hence power law velocity distributions
are  ruled out from the start. 
We then insert Eq. (\ref{eqlv05}) in Eq. (\ref{eq20}) to get
Eq. (\ref{eqlv01}). 
Contrary to the small $\overline{v}$ behavior of the infinite density,
the large $\overline{v}$ behavior is directly related to the
velocity  PDF,
 $F(v)$. 

 In hindsight the asymptotic behavior Eq. (\ref{eqlv01}) 
can be rationalized using a simple argument: 
The large \emph{$2$n}-th order moments $\langle x^{2 n } (t) \rangle$ are
expected to depend only on the rare fluctuations, namely on the 
large $\overline{v}$ limit of ${\cal I}(\overline{v} )$.
With the definition of the infinite density we have
 $P_A(x,t) \sim {\cal I}( x/t)/t^{\alpha}$.
We assume for now that we may  replace $P_A(x,t)$ with
the density $P(x,t)$ and
get $P(x,t) \sim {\cal I}( x/t)/t^{\alpha}$ (see reasoning for this
below).
In this case the moments 
are 
\begin{equation}
\begin{array}{c}
\langle x^{2 n } (t) \rangle = \\
\ \\
 \int_{-\infty} ^\infty P(x,t) 
x^{2 n} {\rm d} x \sim \\
\ \\
\int_{-\infty} ^\infty {{\cal I} \left( { x \over t} \right)\over t^\alpha}  x^{2 n}  
{\rm d} x = \\
\ \\
t^{2 n + 1 -\alpha} \int_{-\infty} ^\infty {\cal I}\left( \overline{v} \right) \overline{v}^{2 n} {\rm d} \overline{v}\:,
\end{array}
\label{eqlv06}
\end{equation}
for $2n \ge 2$. For $n=0$ 
the integral yields infinity as mentioned since 
${\cal I}(\overline{v})$ is not
normalizable.  
Inserting the asymptotic approximation 
Eq. (\ref{eqlv01}) we find by use of an integration by parts
\begin{equation}
\begin{array}{c}
\langle x^{2 n } (t) \rangle \sim\\
\ \\
t^{ 2 n + 1- \alpha} 2 B \int_0 ^\infty { 1 - Q(\overline{v} ) \over \overline{v}  } \overline{v}^{ 2 n}  {\rm d} \overline{v} = \\
\ \\
B 
{ 1 \over 2 n} \langle v^{2 n} \rangle
 t^{2 n + 1-\alpha}. 
\end{array}
\end{equation}
This is indeed the exact result Eq. (\ref{eq01}) in the limit $2 n \gg \alpha$.  
Hence, the large $\overline{v}$ behavior of the ${\cal I}(\overline{v})$,
Eq. (\ref{eqlv01}), 
yields the expected results for the high order moments of
the  spatial displacement. 
We will soon justify the replacement $t^{\alpha} P_A(x,t)$ with $t^\alpha P(x,t)$
which led us to the correct result but first we turn to a few
examples for the infinite density.

\subsection{Examples}

{\it (i)} For a two state model
\begin{equation}
F(v) = \left[ \delta(v-v_0) + \delta ( v + v_0) \right]/2 
\label{eq22}
\end{equation}
we find using Eqs. (\ref{eq20},\ref{eq21})
\begin{equation}
{\cal I}\left(  \overline{v} \right) = \left\{
\begin{array}{ c  c} 
{ B \over 2} \left[ { \alpha (v_0)^\alpha \over |\overline{v}|^{ 1 + \alpha} } -
   { (\alpha - 1) (v_0)^{\alpha -1}  \over |\overline{v}|^\alpha} \right] & |\overline{v}| < v_0 \\
\ & \ \\
0 & |\overline{v}| > v_0 .
\end{array}
\right.
\label{eq23} 
\end{equation}
The particle cannot travel faster than $v_0$ hence the infinite
 density is cutoff
beyond $v_0$ and similarly $P(x,t)=0$ beyond $v_0 t$. 

{\it (ii)}  For an exponential velocity PDF $F(v) = \exp(-|v|)/2$ 
\begin{equation}
{\cal I}(\overline{v}) =
 { B \over 2} \left[ { \alpha \Gamma\left( 1 + \alpha , |\overline{v}|\right) \over |\overline{v}|^{1 + \alpha} } - { \left(\alpha -1\right) \Gamma\left( \alpha , |\overline{v}|\right) \over |\overline{v}|^\alpha} \right] \:, 
\label{eq24}
\end{equation}
and  $\Gamma(a,y)=\int_y ^\infty \exp(-t) t^{a-1} {\rm d} t$ is the incomplete
Gamma function. 

\begin{figure}
\includegraphics[width=0.5\textwidth]{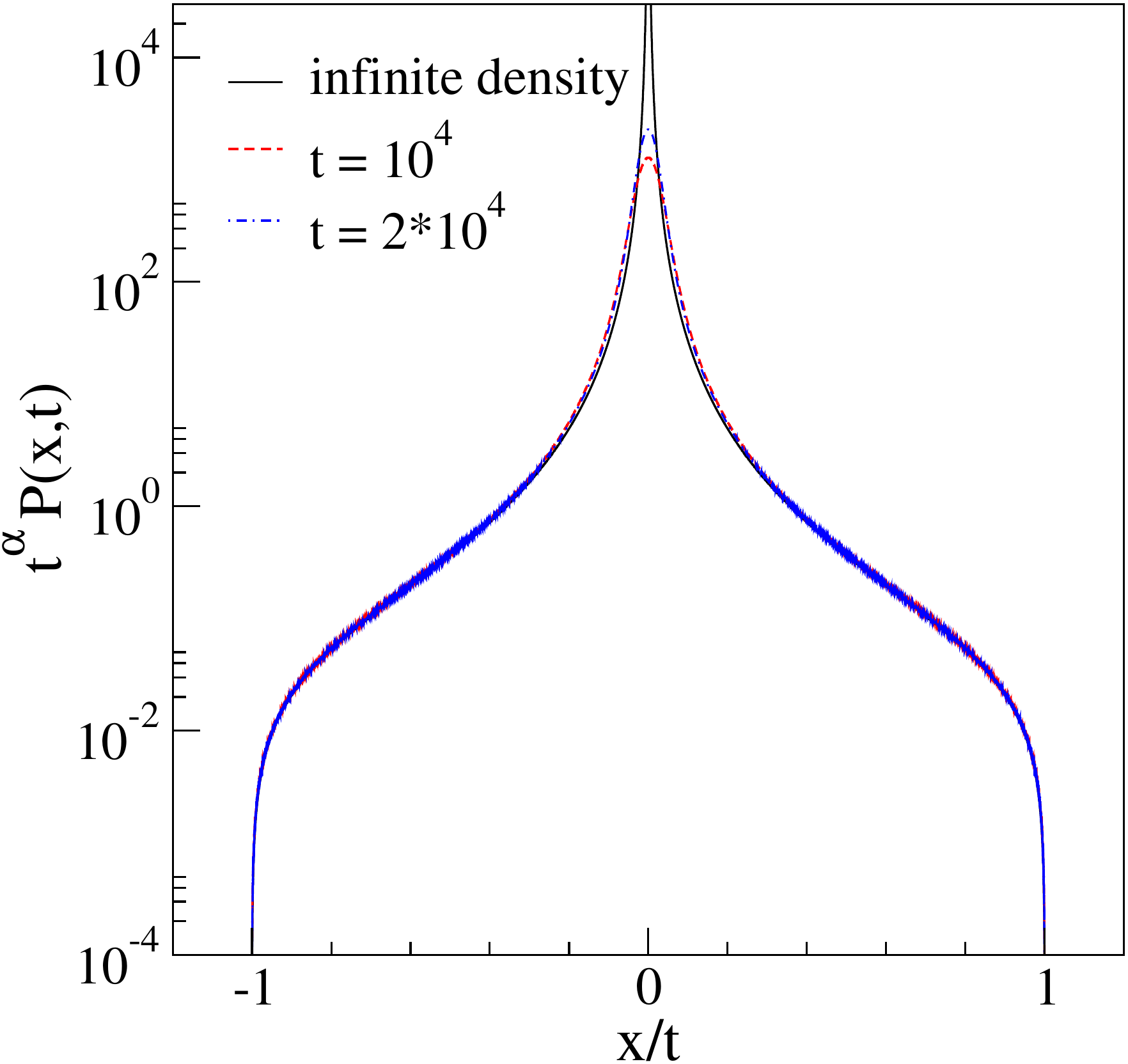}
\caption{(color online) $t^\alpha P(x,t)$ obtained
from numerical simulations  for the model with uniform velocity distribution, $v \in [-1,1]$.
In the long time limit simulations
converge to the infinite density Eq. 
(\ref{eq26}). We used function $\psi(\tau) = \alpha \tau^{ - (1 + \alpha)}$ for $\tau > 1$ as the waiting time PDF.
 The parameters are $\alpha = 3/2$, $A = \alpha |\Gamma(-\alpha)|$, and $\langle \tau \rangle  = 1$.
Two histograms for $t=10^4$ and $2 \times 10^4$
 were  sampled over $N = 10^{10}$ realizations.}\label{Fig1}
\label{fig1}
\end{figure}

\begin{figure}
\includegraphics[width=0.5\textwidth]{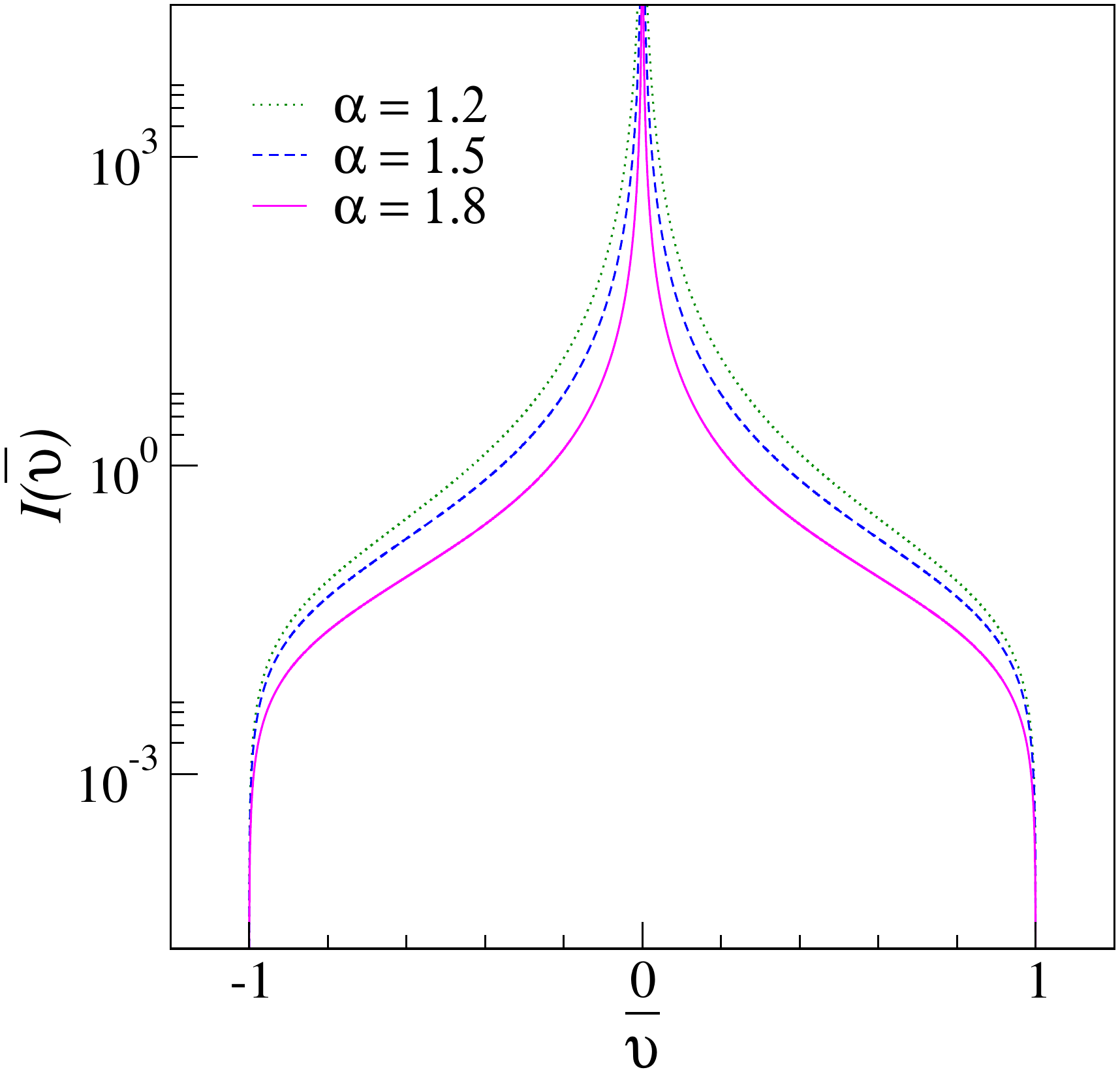}
\caption{(color online) Infinite densities for the model with uniform velocity distribution,
$v \in [-1,1]$, Eq. (55) in the main text. $A = \alpha |\Gamma(-\alpha)|$ and $\langle \tau \rangle  = 1$.}\label{Fig2}
\end{figure}

{\it (iii)}
For a Gaussian model $F(v) = \exp[ - v^2 /2] / \sqrt{ 2 \pi} $ 
\begin{widetext}
\begin{equation}
{\cal  I}(\overline{v}) =
{B \over 2 \sqrt{\pi}} 
\left[ { \alpha \sqrt{2}^\alpha \Gamma\left( { 1 + \alpha \over 2} , { \overline{v}^2 \over 2} \right) \over |\overline{v}|^{ 1 + \alpha}} - { \left( \alpha -1\right) \sqrt{2}^{\alpha -1} \Gamma\left( { \alpha \over 2}, {\overline{v}^2 \over 2}\right) \over |\overline{v}|^\alpha } \right]\:.
\label{eq25}
\end{equation}

{\it (iv)} While for a uniform model $F(v) = 1/2$ for $ -1< v<1$ and
 otherwise
$F(v)=0$ we obtain
\begin{equation}
{\cal I} \left( \overline{v} \right) =
{ B \over 2} 
\left[ { \alpha \over 1 + \alpha} { \left( 1 - |\overline{v}|^{ 1 + \alpha} \right) \over 
 |\overline{v}|^{1 + \alpha} } -
{  \alpha - 1 \over \alpha} { \left( 1 - |\overline{v}|^\alpha \right) \over
|\overline{v}|^\alpha } \right]
\label{eq26}
\end{equation}
for $|\overline{v}|<1$. Similarly to the two state model, the infinite
density 
${\cal I}(\overline{v})$ for the uniform model is zero for $|\overline{v}|>1$, since the particle
cannot travel with a velocity greater than unity (using the correct units). 
In Fig. \ref{Fig1}-\ref{Fig3} infinite densities are plotted
for several values of $\alpha$ and for different models. 

\section{Relation between the density $P(x,t)$ and the infinite density}

We now verify that
 the infinite density yields 
the moments $\langle x^{2 n} (t) \rangle$ with $n=1,2,...$.
Using 
\begin{equation}
\langle x^{2 n} (t) \rangle \sim t^{2 n + 1 - \alpha}  \int_{-\infty} ^\infty
\overline{v}^{2 n} 
{\cal I} \left( \overline{v} \right) 
{\rm d} \overline{v}
\label{eqrel01}
\end{equation}
and  Eq. 
(\ref{eq20}) 
\begin{equation}
\begin{array}{c}
\langle x^{2 n} (t) \rangle \sim 2 B t^{2 n + 1 -\alpha}\int_0 ^\infty \overline{v}^{2 n} \left[ { \alpha {\cal F}_\alpha \left( \overline{v} \right) \over \overline{v}^{1 + \alpha} } - { \left( \alpha - 1 \right) {\cal F}_{\alpha -1} \left( \overline{v} \right) \over \overline{v}^\alpha} \right] {\rm d} \overline{v} = \\
\ \\
2 B  t^{ 2 n + 1 - \alpha}
\int_0 ^\infty \left[ \alpha {\cal F}_\alpha \left( \overline{v}\right) { {\rm d}
 \over {\rm d} \overline{v} }   { \overline{v}^{2 n - \alpha} \over 2 n -\alpha} 
- \left( \alpha - 1\right) {\cal F}_{\alpha -1} \left( \overline{v} \right) {{\rm d} \over {\rm d} \overline{v}}  {\overline{v}^{ 2 n -\alpha +1 } \over 2 n -\alpha +1} \right] {\rm d} \overline{v}\:.
\end{array}
\label{eqrel02}
\end{equation}
Integrating by parts and using Eq. (\ref{eq21}) we find 
the moments, Eq. (\ref{eq01}). Of-course this is  the expected result since we
have constructed the infinite density with the long time behavior of the even
 moments.
\end{widetext}

 We can, in principle, calculate the moments also from the normalized
density because by definition
\begin{equation}
\langle x^{2 n} (t) \rangle = \int_{-\infty} ^\infty x^{ 2 n } P(x,t) {\rm d} x
\:. 
\label{eqrel03}
\end{equation}
The moments in Eq. (\ref{eqrel01}) and Eq. (\ref{eqrel03}) are 
identical in
the long time limit, indicating that the infinite density
and the density $P(x,t)$ are related.  Rewriting Eq. (\ref{eqrel01})
with the change of variables $x=\overline{v}t$
\begin{equation}
\langle x^{2 n}(t) \rangle \sim t^{-\alpha} \int_{-\infty} ^\infty x^{2 n} {\cal I} \left( { x \over t} \right)  {\rm d} x
\label{eqrel03a}
\end{equation}
we conclude from comparison to Eq. (\ref{eqrel03}) 
that
\begin{equation}
{\cal I} \left( {x \over t} \right) \sim { t^\alpha  P (x,t)}
\label{eqrel04}
\end{equation} 
for $x \neq 0$. 
 Thus, we can use the density $P(x,t)$,
 obtained from a finite
time experiment or simulation, to estimate with it the 
infinite density. In the limit $t \to \infty$ the normalized
density multiplied  by $t^\alpha$ and plotted versus $x/t$
 yields  the infinite density. Hence, the infinite
density is not only a  mathematical construction
with which we may attain moments,
rather it contains also information on the positions of particles in space
and hence presents  physical reality in the sense that it is experimentally
  measurable.
We now demonstrate this important observation
 with finite time simulations.

\subsection{Graphical Examples}

 In Fig. \ref{Fig1} we plot $t^\alpha P(x,t)$ obtained from finite time
simulations of the L\'evy walk, versus $x/t$. According to the theory,
in the limit $t \to \infty$ this plot will approach the infinite
density, Eq. 
(\ref{eq20}).
Such a behavior is indeed confirmed with the simulations in the figure.
This implies that with a finite time simulation or experiment, which measures
the density of particles, we can estimate the infinite density.
 In the figure we see that data collapse (for finite time)
is performing better for large $x/t$, and convergence for small
values of $x/t$ is slow. Indeed, for $x=0$ the density is well described
by the L\'evy central limit theorem, as we soon explain. 

\begin{widetext}

\begin{figure}
\includegraphics[width=0.9\textwidth]{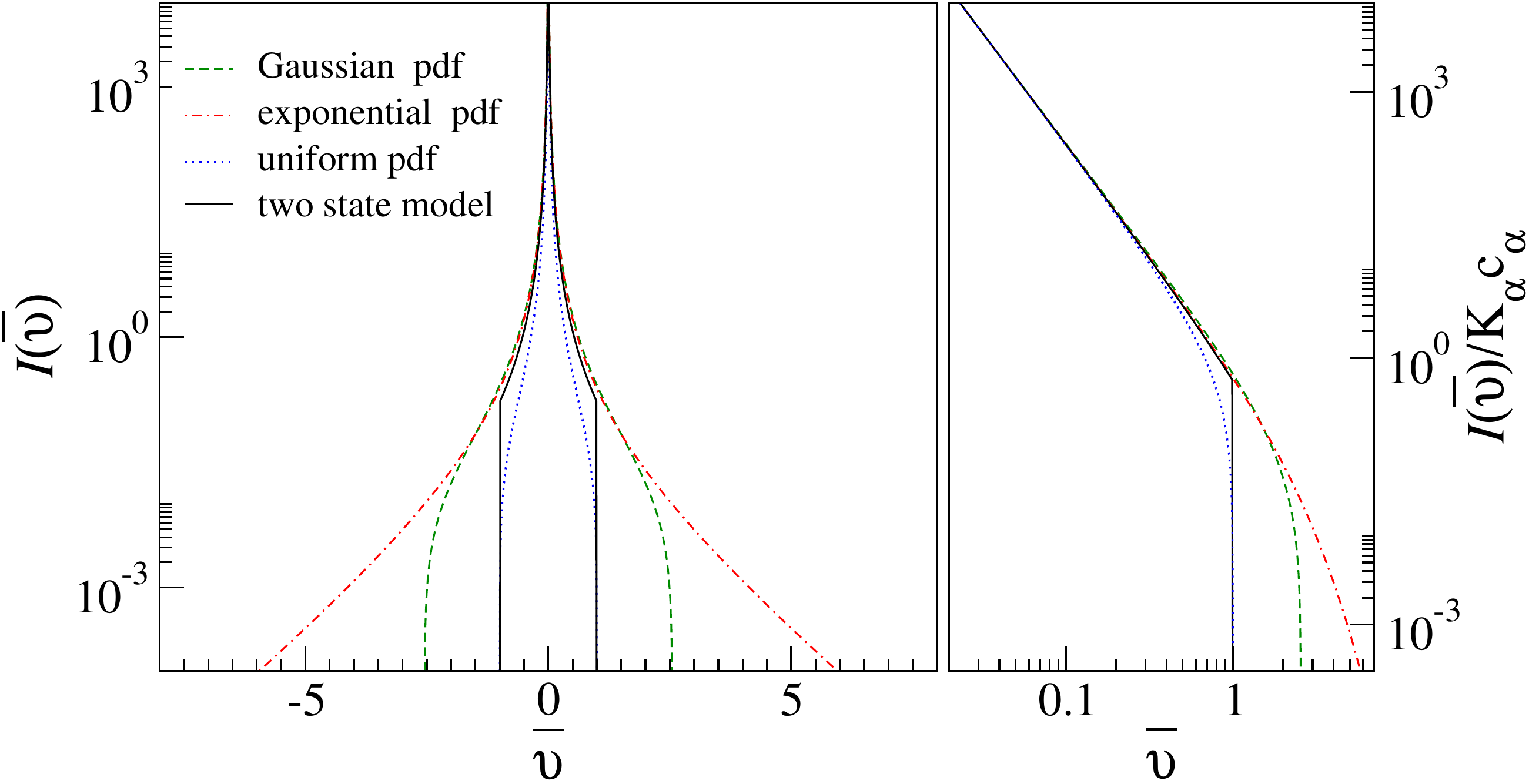}
\caption{(color online) 
Infinite densities for different models, see section ``Examples'' in the main text.
The parameters are the same as in Fig. \ref{Fig1}.
The right panel shows universal behavior
of the infinite density   in the small $\overline{v}$ limit
according to Eq. 
(\ref{eqAs04}).
When the infinite density is rescaled
 with the anomalous diffusion constant $K_\alpha$ times
the constant $c_\alpha$ these
non-normalized densities fall on a master curve. 
Hence once we record  $K_\alpha$ and $\alpha$ we can predict
the small $\overline{v}$ behavior of ${\cal I}(\overline{v})$. 
Here $\alpha=3/2$ and $K_{3/2} =  (\sqrt{2\pi}/3) \langle |v|^{3/2} \rangle$
and $\langle |v|^{3/2} \rangle = 1,2/5, \sqrt{\pi}/4, 2^{3/4} / \sqrt{\pi}$
for the two state, uniform, exponential and Gaussian models respectively.
Notice the log-log scale on the right panel. }
\label{Fig3}
\end{figure}

\end{widetext}

In the right panel in
 Fig. \ref{Fig3} we demonstrate numerically the relation between
the small $\overline{v}$ behavior of ${\cal I}(\overline{v})$
and the anomalous diffusion coefficient $K_\alpha$, Eq. (\ref{eqAs03}).
Plotting ${\cal I}(\overline{v})/ c_\alpha K_\alpha$ versus
$\overline{v}$ we see that
in the limit of small $\overline{v}$,
 various models collapse on one
master curve. 
So with the knowledge of $K_\alpha$ and $\alpha$, which as explained
in the next section can be determined  from data,
 we can predict the small
$\overline{v}$ behavior of the infinite density.  
As mentioned before, the large $\overline{v}$ behavior of the infinite density,  Eq. (\ref{eqlv01}), 
is sensitive to the shape of the velocity CDF $Q(v)$, hence for large $\overline{v}$ 
the corresponding curves in the figure depart. This sensitive tool in real experiments may unravel 
important statistical information on the process.

Practically, though dealing with long time limit solutions, Eq. (\ref{eqrel04}) 
implies one should analyze the results for the shortest valid measurement times 
possible to maximize the probability of the region of interest. 
Thus measurement time must be long, so that asymptotic limit is reached, 
but not too long such that we can sample the tails of the PDF 
(similar to many other problems in physics, for example large deviation theory). 
The estimation of this time scale depends not only on the details of the model, 
but  also on the number of particles undergoing the super-diffusive process, 
and the detailed problem of estimation of infinite densities from finite amount of data, 
is left for future work.

In contrast, the center part of the density,
discussed in the next section, is described by the generalized central limit
theorem, thus it is universal, but does not yield information
on $F(v)$. Thus, infinite densities might become important
tools in unraveling the origin of anomalous diffusion, a
topic which attracted considerable theoretical attention, e.g.,
\cite{PCCP,Garini,Soko} and ref. therein.

In simulations we reached a high degree of accuracy with extensive
simulations and  resolved the probabilities of the events at the PDF tails that are six order of magnitudes smaller than of
those contributing to the PDF maximum, see the right panel of Fig.~1.
The PDFs 
have been sampled with $N = 10^{10}$ realizations.
The corresponding simulations 
were performed on two GPU clusters 
(each consisting of six $\rm{TESLA~K20XM}$ cards)
and took $380$ hours. The main reason 
that we choose such a large number of particles, 
is to illustrate the theory with maximal possible (at the moment) precision. 
Single particle experiments in a single cell \cite{Naama} are conducted with much 
smaller number of particles, and for that reason it would be interesting 
to simulate the process for  smaller numbers of particles and shorter times, 
to see if how accurately one may estimate the infinite density. 
However, in other experiments, like laser cooling of cold atoms  \cite{Sagi}, 
the number of particles is vast so these limitations are absent. 

{\bf Remark:} Upon inspection of Fig. 3, we realize that the 
probabilities of the events contributing to the PDF tails are 
six order of magnitude smaller than those contributing to 
the plotted PDF maximum. This six order of magnitude difference is related to 
the  choice of scale in the figure, since the infinite density diverges on the origin. 
For example in the left panel we cut off the divergence on the origin. 
In reality, experiments are performed for finite times. 
Hence the infinite density is slowly approached, but never actually reached, 
in the vicinity of the origin (see Fig. 1). 
Put differently, we need to sample rare fluctuations, and here sophisticated sampling algorithms could 
be useful \cite{sampling1,sampling2,sampling3}.

%
%
%
%
\section{The L\'evy scaling regime}
\label{SecLregime}

The key step in our derivation of 
${\cal I} \left( \overline{v} \right)$
 was an expansion of the Montroll-Weiss equation, Eq. (\ref{eqLW04}),
in the small $k,u$ limit while the ratio of $k/u$ is kept fixed.
This led to a scaling of the type $x \propto t$. As we showed,
such an expansion leads to a non-normalized
density ${\cal I}(\overline{v})$, as opposed to the normalization dictated by Eq. (\ref{eqLW04}). 

To complete the the large order moments scheme we now consider the expansion of the Montroll-Weiss 
equation assuming that   that $k^\alpha/u$ is fixed and  $u$ is small, i.e.,
we consider the long time limit and inspect the PDF around the origin. Such an assumption means that we are
looking for a solution with a scaling $x \propto t^{1/\alpha}$. This
scaling describes the center part of the PDF $P(x,t)$ \textit{only} and does address the PDF's tails.
It yields a normalized solution but at a price of a divergent second and all higher moments. 
Thus, if we wish to calculate the 
second moment $\langle x^2 \rangle$, we need to estimate the tails of the $P(x,t)$ and use
the infinite density, i.e. perform the fixed $k/u$ expansion.
In contrast, if we want to estimate the normalization, or
the low order moments $\langle |x|^q  \rangle$ with $q<\alpha$,
we need an estimation of the center part of the PDF, and hence and
expansion with $k^\alpha/u$ fixed. Such low order moments
cannot be estimated with the infinite density because
\begin{equation}
\langle |x|^q \rangle = 2  \int_0^\infty P(x,t) x^q {\rm d} x \neq
2 \int_0 ^\infty x^q t^{-\alpha}
 {\cal I} \left( { x \over t} \right) {\rm d} x = \infty 
\label{eqcen01}
\end{equation} 
for $0<q<\alpha$, due to the singular behavior of the infinite density 
at the origin. 

To illustrate the above consideration,  we take Eq. (\ref{eqM02}) and expand it assuming that $|u|\ll |k v| \ll 1$ and $k$ is small,
\begin{widetext}
\begin{equation}
P_{\rm cen} (k,u) \sim \frac{1-A_\tau   \left\langle (- i k v)^{\alpha-1} \left[1-(\alpha-1)\frac{u}{i k v} \right]\right\rangle}{u-A_\tau   \left\langle (- i k v)^{\alpha} \left[1-\alpha\frac{u}{i k v} \right]\right\rangle} \sim
\frac{1}{u-A_\tau   \left\langle (- i k v)^{\alpha}\right\rangle}.
\label{p_center_ku11}
\end{equation}
\end{widetext}
The subscript ``${\rm cen}$" indicates we aim at the center 
part of $P(x,t)$. 
The term $ \left\langle (- i k v)^{\alpha}\right\rangle$ can further be simplified  by recalling the symmetry of $F(v)$, 
\begin{equation}
\begin{array}{rl}
 \left\langle (- i k v)^{\alpha} \right\rangle &=|k|^\alpha \left\langle |v|^{\alpha} e^{-\frac{i \pi \alpha }{2} \text{ sign}(k v)} \right\rangle\\
\\
&=|k|^\alpha \left\langle |v|^{\alpha} \right\rangle \cos{\left( \frac{\pi \alpha}{2} \right)}.
\end{array}
\label{p_center_ku_aid}
\end{equation}
By inserting it into Eq. (\ref{p_center_ku11}) we find
\begin{equation}
P_{\rm cen} (k,u)  \sim  \frac{1}{u+K_{\alpha}|k|^\alpha } ,
\label{eqcen02}
\end{equation}
where $K_\alpha$ is the anomalous diffusion coefficient, proportional to the moment
$\langle |v|^\alpha\rangle$  of  the PDF $F(v)$, Eq. (\ref{eqAs03}).
Thus, here the dynamics is not sensitive to the full shape of the
velocity distribution but only to a  particular $\alpha$-th moment. 

The solution Eq. (\ref{eqcen02}) is well known 
\cite{Bouchaud,GL,Review,ZK,Carry}; its
inverse  Laplace transform is 
\begin{equation}
P_{{\rm cen}}(k,t) \sim \exp \left( - K_\alpha t |k|^\alpha\right)\:. 
\label{eqcen05}
\end{equation}
As is well known, the Fourier transform of the symmetrical
L\'evy density  $L_\alpha(y)$ is $\exp(-|k|^\alpha)$,
which serves as our working definition of this stable density. 
Hence by definition, the inverse Fourier transform
 of  Eq. (\ref{eqcen05})
yields a symmetric L\'evy
stable PDF \cite{Bouchaud,Review,levy} 
\begin{equation}
P_{\rm cen}(x,t) \sim{ 1 \over (K_\alpha t)^{1 /\alpha}}
 L_\alpha\left[ { x \over (K_\alpha t)^{1/\alpha}} \right]\:.
\label{equation}
\end{equation}
The L\'evy density is normalized, its second moment diverges
 since for large $x$ the solution has
a fat  tail $L_\alpha(x)\propto|x|^{-(1 + \alpha)}$. 
When $\alpha \to 2$ we approach the Gaussian limit. 
\begin{figure}[t]
\includegraphics[width=0.5\textwidth]{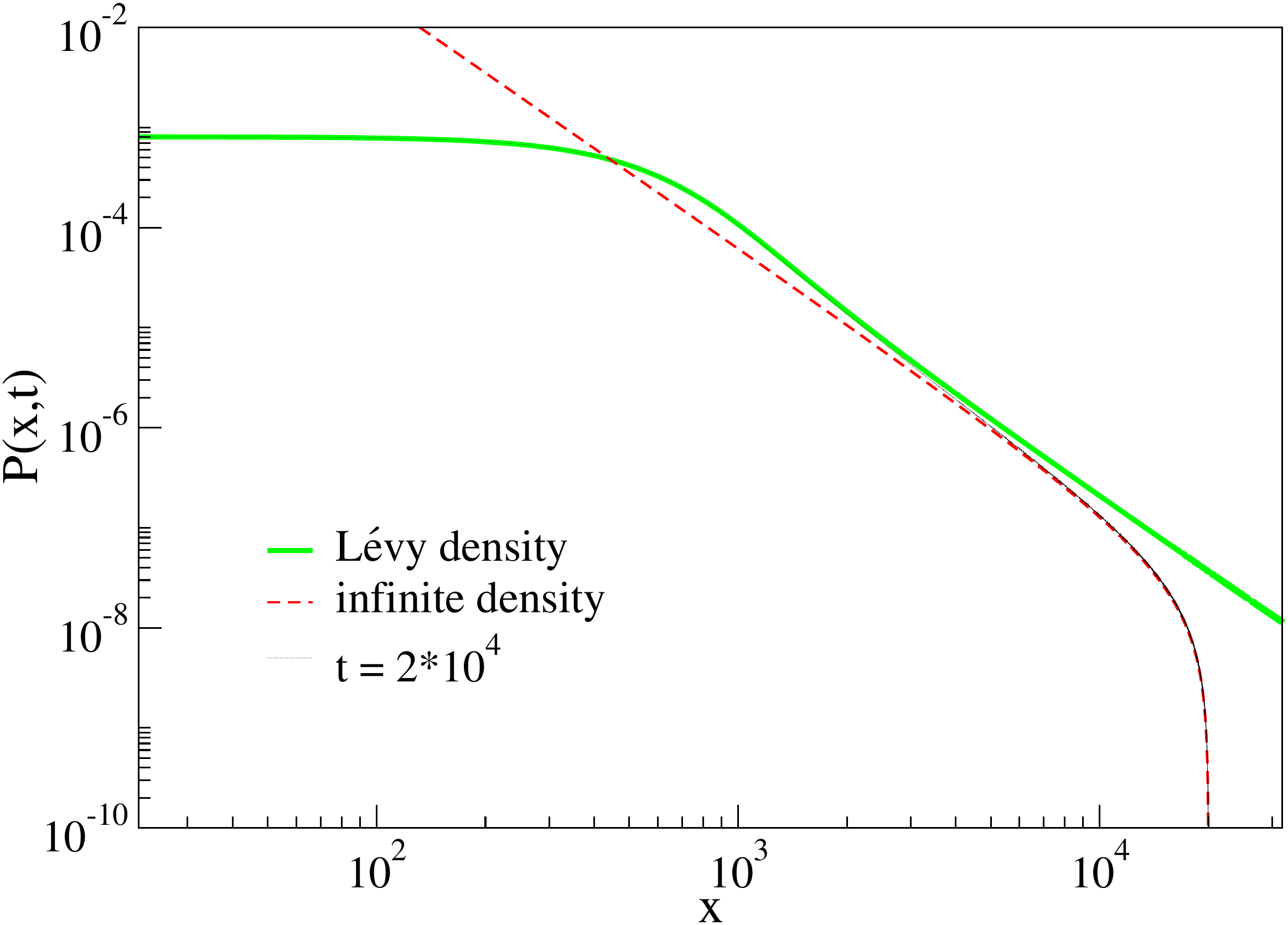}
\caption{(color online) PDF $P(x,t)$  for the
model with uniform velocity distribution, $v \in [-1,1]$, see Fig. \ref{Fig1},
 plotted together with
L\'{e}vy distribution and rescaled infinite density. The parameters are the same as in Fig. \ref{Fig1}.}\label{Fig4}
\end{figure}

One cannot say that either of the two Montroll-Weiss expansions presented so far is the correct expansion, without specifying the
observable of interest and the domain in $x$ where one wishes
to estimate the density.  Thus, if we wish to calculate the 
second moment $\langle x^2 \rangle$ from an estimation of the density
of particles, we need the infinite density (i.e. the 
fixed $k/u$ expansion) estimating the tails of the $P(x,t)$.
In contrast, if we want to estimate the normalization, or
the low order moments $\langle |x|^q  \rangle$ with $q<\alpha$,
we need an estimation of the center part of the packet, Eq. (\ref{equation}).

\section{Infinite and L\'evy densities are complementary}

 As shown in Fig. \ref{Fig4}, for long, though finite times 
the center/tail region of $P(x,t)$ is well approximated by the L\'evy/infinite
density respectively. 
We define a crossover position
$x_c(t)>0$ below/above which the L\'evy/infinite densities are valid
approximations. The maximum of $P(x,t)$ is clearly on the origin
(neglecting possible finite time delta peaks found, for example,
for the two state velocity model,
since  these  do not contribute
in the $t\to \infty$ limit to the moments of the process). 
Hence, any estimation for the density for $x\neq 0$
and long times must be smaller than the value of the density on the origin. 
On the origin
\begin{equation}
P(x,t)|_{x=0} \sim { L_\alpha(0)  \over (K_\alpha t)^{1/\alpha}}\:,
\label{secCom01}
\end{equation}
with  $L_\alpha(0) = \int_{-\infty} ^\infty \exp( - |k|^\alpha) {\rm d} k/ 2 \pi = \Gamma(1 + \alpha^{-1})/\pi$.
Using $P(x,t)\sim {\cal I} (x/t) /t^{\alpha}$ we define $x_c$ via 
the condition
\begin{equation}
{ {\cal I} \left( { x_c \over t} \right) \over t^{\alpha}} = { L_\alpha\left( 0 \right) \over \left( K_\alpha t \right)^{1/\alpha} } \:.
\label{secCom02}
\end{equation}
Since the transition to the L\'evy regime is found for the small
argument behavior of the infinite density we
use Eq. (\ref{eqAs04}),
$ {\cal I} ( x_c/t) \sim c_\alpha K_\alpha / |x_c/t|^{1 + \alpha}$
and after some simple algebra we find
\begin{equation}
x_c(t) = \left[ { c_\alpha \over L_\alpha(0)} \right]^{{1 \over 1 +\alpha}}
\left( K_\alpha t\right)^{1/\alpha}\:.
\label{secCom02}
\end{equation}
Though only
 a rough estimation for the transition, the important point is to notice
that $x_c (t) \propto t^{1/\alpha}$. When data for $P(x,t)$
 is presented versus the scaled variable 
$x/t$ (the variable $\overline{v}$) we have a crossover velocity $x_c(t)/t\sim
t^{(1/\alpha)-1}$ which goes to zero as $t \to \infty$ since $\alpha>1$.
  Thus, the
infinite density becomes a better
approximation of  $t^\alpha P(x,t)$ versus $x/t$ 
as time is increased, see Fig.  \ref{Fig1}. 

\begin{widetext}
 Based on this picture let us calculate the $q$-th
 moment with $q>\alpha$, being a real number. We divide the spatial integration into two parts and find
\begin{equation} 
\langle |x|^q \rangle \simeq
\underbrace{ 2 \int_0 ^{x_c(t)} {L_\alpha \left[ {x \over \left( K_\alpha t \right)^{1/\alpha}} \right] \over \left( K_\alpha t\right)^{1/\alpha} } x^q {\rm d} x  }_{\text{ 
\mbox{inner L\'evy region}}}
+ \underbrace{ 2 \int_{ x_c(t) } ^\infty { {\cal I} \left( {x \over t} \right) \over t^\alpha} x^q {\rm d} x}_{\text{ outer tail, infinite density region}}\:. 
\label{secCom03}
\end{equation}
Changing variables according to $y=x/(K_\alpha t)^{1/\alpha}$ in the
first integral on the right hand side, and to $\overline{v} = x/t$ 
for the second integral
we have
\begin{equation}
\langle |x|^q \rangle \simeq 2 (K_\alpha t)^{q/\alpha}
 \int_0 ^{x_c(t)/(K_\alpha t)^{1/\alpha} }
L_\alpha(y) y^q {\rm d} y
+ 2 t^{ q + 1 -\alpha}   \int_{x_c(t)/t} ^\infty {\cal I} \left( \overline{v} \right) \overline{v}^q {\rm d } \overline{v}\:.
\label{secCom04}
\end{equation} 
In the long time limit, the lower limit of the second integral $x_c(t)/t\to 0$
while the upper limit of the first integral is a constant.
For
 $q>\alpha$ the second integral is by far larger than the first, hence we may neglect
the inner region 
\begin{equation}
\langle |x|^q \rangle \sim 2 t^{ q + 1 -\alpha} \int_0 ^\infty {\cal I} \left( \overline{v} \right) \overline{v}^q {\rm d} \overline{v}.
\label{secCom05}
\end{equation} 
The $\overline{v}^q$ in the integrand
``{\it cures}'' the non-integrable infinite density, i.e.,
the non-integrability arising from
the small $\overline{v}$ divergence  of ${\cal I}(\overline{v})$, in the sense
that the integral is finite when $q>\alpha$.
When $q=2 n$ in Eq. (\ref{secCom05}) with a positive integer $n$
we retrive  Eqs. (\ref{eq01},\ref{eqrel03a}). 

In the opposite limit $q<\alpha$ we find the opposite trend, namely
now the inner region of the density $P(x,t)$ is important in the estimation
of the moments $\langle |x|^q \rangle$. To see this we first note that
in the intermediate region of $x$, the L\'evy and infinite densities
are equivalent. The intermediate region are values of $x$ where the density
$P(x,t)$ is well approximated by the power law tail of the L\'evy density,
before the cutoff due the finite velocity, and after the
small $x$ region where the L\'evy density did not yet settle into
a power law behavior. Indeed, there  is a relation between
the infinite density and the L\'evy density since they must match
in the intermediate region. Using
the known large $y$
 behavior  $L_\alpha(y) \sim c_\alpha y^{ -(1 + \alpha)}$
we obtain
\begin{equation} 
t^{\alpha} P_{{\rm cen}} (x,t) \sim c_\alpha K_\alpha |x/t|^{-(1 +\alpha)}\:.
\label{eqComaa}
\end{equation}
This is exactly the same as the small $\overline{v} = x/t$ behavior of
the infinite density, 
Eq. (\ref{eqAs04}), hence, the two solutions match as they should. 

To obtain $\langle |x|^q \rangle$ with $q<\alpha$ we  
  define a velocity  crossover $v_c$ which is of the order
of the typical velocity of the problem (a velocity scale of $F(v)$).
For $|x|/t<v_c$ the density $P(x,t)$ is well
approximated by a L\'evy density (since in the intermediate region the
latter and the infinite density  match), 
while beyond $v_c$ we have the 
infinite density description
\begin{equation}
\langle |x|^q \rangle \simeq 2 \int_0 ^{v_c t} { L_\alpha \left[ { x \over \left( K_\alpha t\right)^{1/\alpha}} \right] \over \left( K_\alpha t \right)^{1/\alpha} }  x^q {\rm d} x +
2 \int_{v_c t} ^\infty 
{ {\cal I} \left( { x \over t} \right) \over t^\alpha} x^q {\rm d} x\:.
\label{secCom06}
\end{equation}
Now when $t \to \infty$ and $q<\alpha$ the contribution from the second integral
is negligible and the upper limit of the first integral is taken
to infinity, hence, after a change of variables
$y = x/ ( K_\alpha t)^{1/\alpha}$, and using
the symmetry $L_\alpha(y) = L_\alpha(-y)$
\begin{equation}
\langle |x|^q \rangle \sim \left( K_\alpha t\right)^{ q \over \alpha} \int_{-\infty} ^\infty L_\alpha(y) |y|^q {\rm d} y\:. 
\label{secCom07}
\end{equation}
 We see that moments integrable with respect to  the infinite
density, i.e. $q>\alpha$ are obtained from the non-normalizable measure.
While for an observable which is non-integrable with respect to the infinite
density, i.e. $|x|^q$ and $q<\alpha$,
the L\'evy density yields the average and is used in the calculation. 

 Further  mathematical analysis of this behavior is required.
An observable like $f(x)=1/(1+\sqrt{|x|})$ is non integrable with respect
to the infinite density. The average
of $f(x)$ with respect to  the L\'evy density is finite.
Hence  the average $\langle f(x) \rangle$
 is computed with an integration over  the L\'evy density.
On the other hand consider an observable like 
$g(x)= |x|^{\alpha + \epsilon}\theta(1-|x|)$ where $\theta(\cdots)$ is
 the step function and $\epsilon > 0$. Now $g(x)$ is integrable
with respect to the infinite density but also integrable with
respect to the L\'evy density. So how do we obtain its average?
Do we use the infinite density or the L\'evy density?
Since $P(x,t)$ is well approximated by the L\'evy density in the
regions where the function $g(x)$ is not zero, i.e. for $|x|<1$
the L\'evy density should be used in the calculation.
This example shows that not all averages of
 observables integrable with respect to the infinite density
are obtained from the non-normalized density. 
One can also construct an observable which is not integrable with
respect to the L\'evy density neither with respect to
the infinite density; e.g., a function that behaves like $h(x) \sim x^2 $ for
large $x$ and $h(x) \sim 1 $ for $x\to 0$. A trivial example is
 $h(x) = 1 + x^2$ since the average of this function is given by 
both  the L\'evy density
to compute  $\langle 1 \rangle=1$ and
 $\langle x^2 \rangle >> 1$ which is found from the infinite density. 
Hopefully future rigorous
 work will present a full classification of observables
and rules for their calculation. 
Another approach it to  obtain a uniform approximation
for $P(x,t)$ based on the L\'evy and infinite densities.
This approximation which works for all $x$
 will be presented elsewhere and it can serve as a tool for
calculation of different classes of observables.

\end{widetext}

\subsection{Strong Anomalous Diffusion}

\begin{figure}[t]
\includegraphics[width=0.5\textwidth]{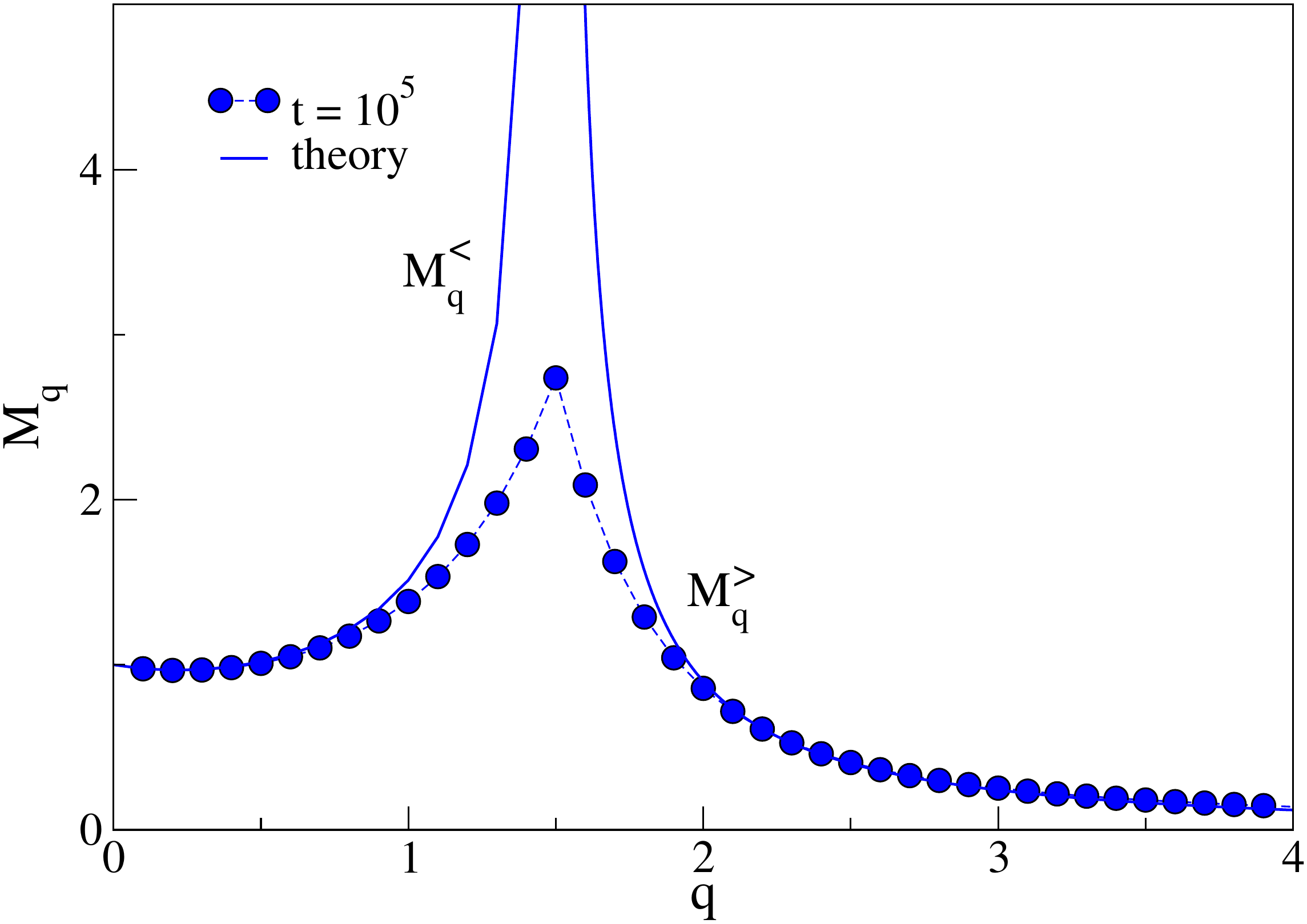}
\caption{(color online) Moment amplitudes for the uniform model 
exhibit a critical behavior on $q=\alpha$, in the infinite long time limit these
moments diverge as $q \to \alpha$ $(\alpha =3/2$ in this example).
 Lines correspond to Eqs. 
(\ref{SecSt02},
\ref{SecSt03})
 in the main text.
Dots are the results of sampling with $N = 10^9$ realizations. Dashed line was obtained by combining
together the L\'{e}vy distribution and infinite density (see Fig. \ref{Fig4}) 
and gluing them together
at the point $x = 15000$ 
\cite{remark1}.
 Parameters are the same as in Fig. \ref{Fig1} with $t=10^5$.
}\label{Fig5}
\end{figure}

\begin{figure}
\includegraphics[width=0.5\textwidth]{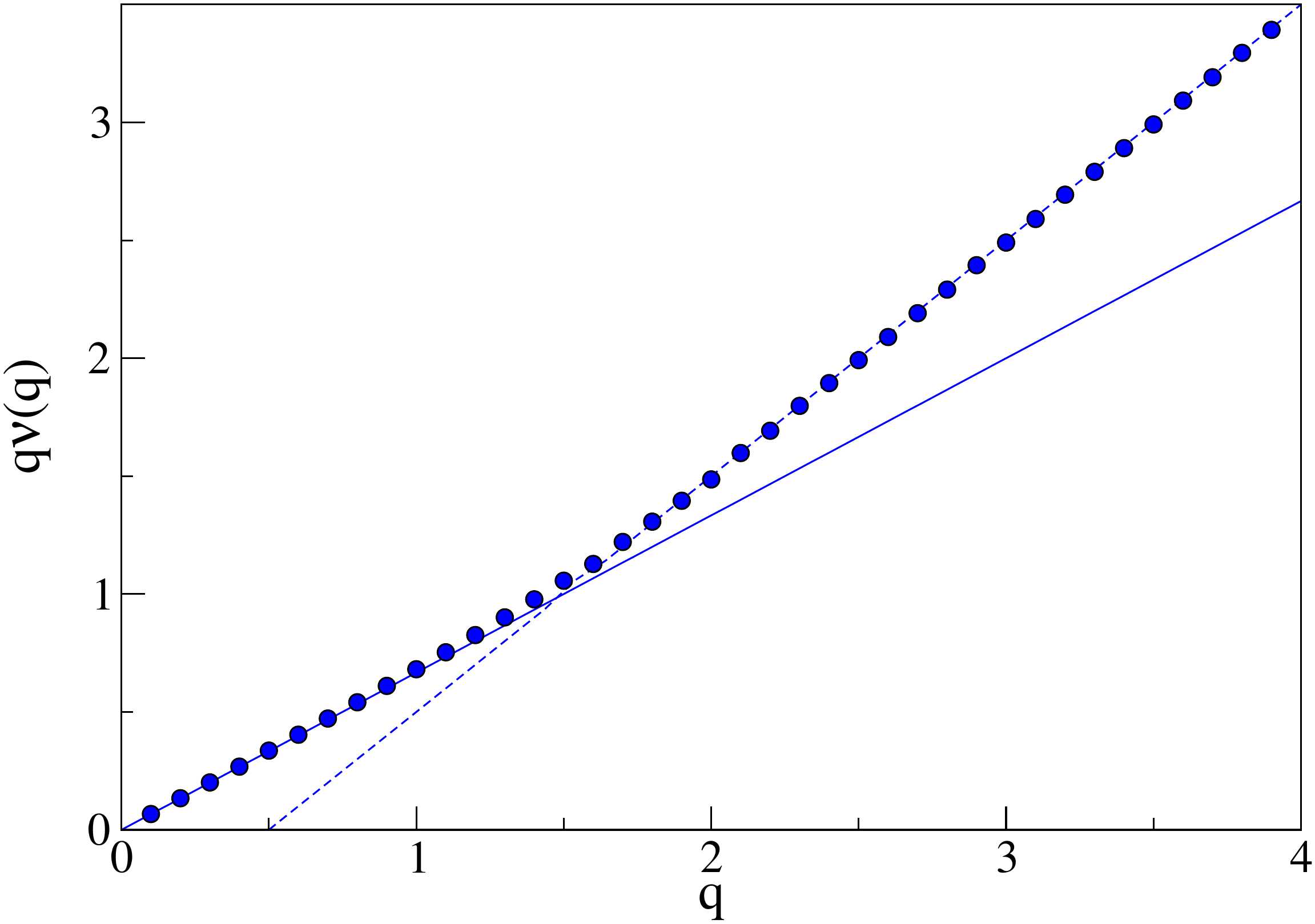}
\caption{(color online) 
The spectrum of exponents $q \nu(q)$ versus $q$ exhibits piecewise
linear behavior. The finite time simulations (dots, parameters as
in Fig. \ref{Fig5}),
 for the uniform  model,
 perfectly agree with
theory (lines): for $q<\alpha$ $q \nu(q)=q/\alpha$ otherwise
$q \nu(q) = q + 1 - \alpha$ (see Eq.
\ref{SecSt01}).
Here $\alpha =3/2$. 
}
\label{Fig6}
\end{figure}

 From Eqs. 
(\ref{secCom05},
\ref{secCom07}) we find
\begin{equation}
\langle |x|^q \rangle \sim \left\{
\begin{array}{l l}
M_{q} ^{<}\ t^{q/\alpha}, \  & q<\alpha, \\
\ & \ \\
M_{q} ^{>}\  t^{q + 1 -\alpha}, \  & q> \alpha,
\end{array}
\label{SecSt01}
\right.
\end{equation}
with amplitudes
\begin{equation}
M_{q} ^{<} = { \left( K_\alpha \right)^{ q \over \alpha} \Gamma\left( 1 - { q \over \alpha} \right) \over \Gamma\left( 1 - q \right) \cos \left( \pi q /2 \right) } 
\label{SecSt02}
\end{equation}
and
\begin{equation}
M_{q} ^{>} = { 2 K_\alpha c_\alpha q \langle |v|^{q} \rangle \over \alpha\left( q -\alpha\right) \left( q - \alpha + 1 \right) \langle |v|^\alpha \rangle }\:. 
\label{SecSt03}
\end{equation}
The amplitudes $M_{q} ^{<}$ and $M_{q} ^{>}$ diverge as $q\to \alpha$ from above
or below, and in that sense the system exhibits a dynamical phase
transition. This behavior is shown in Fig. \ref{Fig5} where finite time
simulations show a clear peak of the moments amplitudes
 in the vicinity of $q=\alpha$.
 The system exhibits strong anomalous diffusion with a bi-linear
spectrum of exponents, i.e., $q \nu(q)$ in Eq. 
(\ref{eqInt01}) is bi-linear. 
Such a behavior is demonstrated in Fig. \ref{Fig6},
where $q \nu(q)$ versus $q$ is plotted
using finite time simulations,
which indicate that convergence to asymptotic results is within reach
\cite{Zabu1}.
As mentioned in the introduction,
 this piecewise linear behavior of $q \nu(q)$,  
is  a
widely observed behavior. 

\section{Discussion}

The L\'evy walk model
exhibits enhanced diffusion where $\langle x^2 \rangle
\propto t^{3-\alpha}$ when $1<\alpha<2$. 
Such a behavior is faster than diffusive and slower
than ballistic.  
Unlike Gaussian processes with zero mean, the variance $\langle x^2 \rangle$ 
is not a sufficient characterization of the motion. 
Mono scaling solutions $P(x,t) \sim t^{-\xi} g(x/t^\xi)$,
 which assume that the density of
particles in the long time limit has a single characteristic scale,
fail to describe the dual nature of the dynamics which contains
both L\'evy motion and ballistic elements in it. 
 The density of particles at its center part is described by the symmetric
L\'evy distribution. 
If the L\'evy central limit theorem
is literally taken, it predicts a divergence of the mean square displacement
at finite times, i.e., the variance of the stable PDF $L_\alpha(x)$ is infinite
when $0<\alpha<2$.
The tails of  the L\'evy walk
 model exhibit ballistic scaling, indeed as well known, 
 the density is cutoff by ballistic flights. In this work we have
found the mathematical description of the ballistic elements of the transport.
An uncommon physical tool,  a non-normalizable
density describes the  packet of particles.
The bi-linear scaling of the moments, i.e., strong anomalous diffusion,
 means that we have two complementary scaling solutions
for the problem. The first describes the center part of the packet and is 
the well known L\'evy stable density
with the scaling $x \propto t^{1/\alpha}$. 
 The second scaling solution
describes ballistic scaling $x \propto t$
and is given by our rather  general formula
 for the infinite density Eq. 
(\ref{eq20}).
The equation exhibits a certain universality in the sense that it
does not depend on the full shape of the waiting times PDF.
It relates the non-normalized density with
three measurable quantities:  the velocity distribution of
the particles,
the anomalous diffusion  exponent $\alpha$ 
and the anomalous diffusion coefficient $K_\alpha$.

 In this paper we have focused on a model with two scalings, namely
ballistic $x \propto t$ and L\'evy $x \propto t^{1/\alpha}$
motions.
 In real systems we may have a mixture of 
other modes of motion. For example, Gal and Weihs \cite{Naama} measured the
spectrum of exponents $q \nu(q)$ and found for large $q$ a linear
behavior $q \nu(q) \sim c q$ and $c\simeq 0.8-0.6$ so this motion
is slower than ballistic but faster than diffusive.
 This is probably related to the active transport
of the measured  polymer embedded in the live cell,
 namely to the input of energy into the cell. 
In this direction, one should consider more general models.
For example variation of the L\'evy walks where waiting times
are power law distributed but jump's lengths scale non-linearly with 
waiting times, i.e. $x = \sum_{i=0}^N v_i (\tau_i)^\beta + v_N (\tau_b)^\beta$
with  $\beta>0,\beta \neq 1$. 
Work on other regime of parameters is also in progress.
We recently investigated the regime $\alpha>2$, where
the Gaussian central limit theorem holds, and found that the infinite density
concept remains in tact. Thus even a normal process, in the sense that
the  center part of $P(x,t)$ 
is described by the Gaussian central limit theorem, its rare
fluctuations are related to  a non-normalizable measure. 
It is left for future rigorous
work to see if infinite densities describe non-linear dynamical
systems, at least those where power law distributions of waiting
times between collision events,  are known to describe the dynamics \cite{Dettman,Balint}.

Our results seem widely applicable. We know that bi-scaling of the spectrum of
exponents is a common feature of different systems and strong anomalous diffusion is 
a well documented phenomena. The L\'evy walk dynamics is ubiquitous and has been recorded 
in many systems. Hence we are convinced  that
the infinite density concept has a general validity,
ranging from dynamics in the cell, motion of tracer particles
in non-linear flows, spatial diffusion of cold atoms
to name a few. The question of estimation of infinite density from 
not too large ensembles of particles and finite time experiments is left for future work. 

 In solid state physics, transport is in many cases characterized
as either ballistic or diffusive. The system falls into one
of these categories depending on the ratio of the size
of the system and mean free path. Here a completely different
picture emerges. Depending on the observable, i.e. the order of the moment,
 the same system exhibits
either  diffusive or ballistic transport. Observables integrable
with respect to the infinite density are ballistic, while observables
integrable with respect to the L\'evy density exhibit super diffusion.
So in these systems the question of what is measured becomes crucial,
and one cannot say that the process/system itself
falls into a unique category
of motion. At-least in our model,
bi-scaling means that we have two sets of tools to master,
the infinite density being the relatively newer concept which might need
more clarifications in future work \cite{Large} and become a valuable approach
in other problems of statistical physics.

\acknowledgments
This  work  was supported by the  Israel Science  Foundation (AR and EB),
the German Excellence 
Initiative ``Nanosystems Initiative Munich''(SD and PH),
and by the grant $\mbox{N}  02.49.21.0003$ (the agreement of
August $27$, $2013$  between the Ministry of Education and Science of the Russian Federation
and Lobachevsky State University of Nizhni Novgorod) (SD).
EB thanks the Alexander von Humboldt foundation for its support.

\section{Appendix:  $P(k,u)$ for the L\'evy walk}

In this Appendix we derive the Montroll-Weiss equation 
(\ref{eqLW04}) relating 
the Fourier-Laplace transform of $P(x,t)$ with the velocity
and waiting time PDFs, $F(v)$ and $\psi(\tau)$ respectively. 
We denote the position of the particle $x_N(t)$ (instead of $x(t)$ used
in the text), where $N$ is the random number
of collisions and  rewrite Eq. 
(\ref{eqModel00})
\begin{equation}
x_N(t)  = \sum_{j=1} ^N v_{j-1} \tau_j + v_N \tau_b\:. 
\label{eqAp01}
\end{equation}
The standard approach to the problem is an iterative approach,
namely  to relate the probability density
 of finding the
particle at $x$ after $j$ collision events, with the
  density  
conditioned on $j+1$ collisions, e.g. \cite{Carry}.
 Using the
renewal assumption
one gets convolution integrals, and that leads to Eq. (\ref{eqLW04}).
We will use a slightly different method, to avoid a complete repeat of 
previous derivations, our approach is inspired by
 the renewal theory in \cite{GL}.

The PDF of the position of the particles, all starting on the origin $x=0$,
 at time $t$ is
\begin{equation}
P (x,t)=\sum_{N=0}^{\infty} {\langle \Theta(t-t_N) \Theta(t_{N+1}-t) \delta[x-x_{N} (t)] \rangle}\:,
\label{eq_pxt_vel}
\end{equation}
where $\delta(\cdots)$ is the Dirac delta function.
Here $t_N$ denotes the time  of the  $N$-th collision event
$t_N = \sum_{i=1} ^N \tau_i$.  
The multiplication of the two  step functions in Eq. (\ref{eq_pxt_vel}),
i.e., the $\Theta(t-t_N) \Theta(t_{N+1}-t)$ term, gives the condition 
$t_{N} < t < t_{N+1}$, for the measurement time $t$.
The summation over $N$ in Eq. (\ref{eq_pxt_vel}) is a sum over
all possible number of collision events. 
Transforming $P(x,t)$ into  the Fourier-Laplace domain, using
Eq. (\ref{eqLW03}), we obtain
\begin{equation}
P (k,u)=\sum_{N=0}^{\infty} {\left\langle \int_{t_N}^{t_{N+1}} {\text{exp}({i k x_{N}  (t) - u t }) {\rm d} t}\right\rangle}\:.
\label{eq_pku_vel1}
\end{equation}
The averages here are with respect to the velocity and 
the waiting time distributions. Inserting Eq. (\ref{eqAp01})
in Eq. (\ref{eq_pku_vel1})
 we perform the time  integral on the right hand side
 of Eq. (\ref{eq_pku_vel1}) using $t_{N+1}-t_{N}=\tau_{N+1}$
and $\tau_b= t-t_N$
\begin{widetext}
\begin{equation}
\begin{array}{rl}
P (k,u)&=\sum_{N=0}^{\infty} {\left\langle \frac{1-\text{exp}[{(i k v_N-u) \tau_{N+1}}]}{u-i k v_N } { \prod_{j=1}^N \text{exp}[{(i k v_{j-1}-u) \tau_j}]} \right\rangle} \\
\\
&=\left\langle \frac{1-\text{exp}[{(i k v_0-u) \tau_{1}}]}{u-i k v_0} \right\rangle+\sum_{N=1}^{\infty} {\left\langle \frac{1-\text{exp}[{(i k v_N-u) \tau_{N+1}}]}{u-i k v_N } \prod_{j=1}^{N} \text{exp}[{ (i k v_{j-1}-u) \tau_j}] \right\rangle}\:.
\end{array}
\label{eq_pku_vel2}
\end{equation}
Here we separated the case of zero collisions $N=0$ from $N\ge 1$.
Because the velocities and waiting times are mutually independent,
each separately being independent identically distributed
random variables, we can perform
the averaging.  With the Laplace
transform
Eq. (\ref{eqMod01}),
 $\hat{\psi}(u)=\left\langle e^{-u \tau }\right\rangle$, 
we use
\begin{equation}
\langle \Pi_{j=1} ^N \exp[(i k v_{j-1} - u) \tau_j]\rangle=
\langle \hat{\psi}( u - i k v) \rangle^N\:,
\label{eqAp03}
\end{equation} 
where the remaining  averaging on the right hand side is with respect
to the velocity PDF only $\langle \cdots \rangle= \int_{-\infty} ^\infty {\rm d} v \cdots F(v)$.
Hence from Eq. (\ref{eq_pku_vel2}) we find
\begin{equation}
P(k,u) = 
\left\langle \frac{1-\text{exp}[{(i k v-u) \tau}]}{u-i k v} \right\rangle \sum_{N=0}^{\infty} { \left\langle \text{exp}[ {(i k v-u) \tau}] \right\rangle^N}=\left\langle \frac{1-\hat{\psi}(u-i k v)}{u-i k v} \right\rangle \sum_{N=0}^{\infty} { \left\langle \hat{\psi}(u-i k v) \right\rangle^N}\:.
\label{eq_pku_vel3}
\end{equation}
\end{widetext}
This geometric series sum is convergent and yields the known Montroll-Weiss equation, Eq. (\ref{eqLW04}).


\begin{thebibliography}{99}
\bibitem{remark1} In Fig. \ref{Fig4} $t=2 \times 10^4$ and hence the
position $x$ where the  two solutions overlap is roughly
$x\simeq 3000$ while in Fig. \ref{Fig5} $t=10^5$ hence this point
is shifted towards $15000$.
\bibitem{Large}
Large deviation theory deals
with remote tails of distribution functions,
for example with the classical problem of summation of independent random
variable with a common PDF.  In this example large deviations theory
contains the Gaussian central limit theorem and extends it to the regime of
large fluctuations \cite{Hugo}.  It is still left to be seen if
large deviations theories can be used to obtain infinite densities
for the L\'evy walk problem.
\end{thebibliography}
\end{document}